\documentclass[pra,twocolumn,floatfix,showpacs]{revtex4}
\usepackage{amsmath,amssymb,amsfonts,bbm,graphicx,hyperref,times,psfrag}
\usepackage{amsthm}
\usepackage{amsfonts}
\usepackage{microtype}
\usepackage{float}
\usepackage{mathrsfs}
\usepackage{placeins}
\usepackage{bbm}
\usepackage{color}
\usepackage{epstopdf}

\newcommand{\be}{\begin{equation}}
\newcommand{\ee}{\end{equation}}

\newcommand{\daga}{^{\dagger}}
\newcommand{\bsigma}{\boldsymbol \sigma}
\newcommand{\bbeta}{\boldsymbol \beta}

\def\sts{\!\hbox{\tiny STS}\,}
\def\es{\hbox{\tiny E}\,}
\def\tc{\hbox{\tiny C}\,}
\def\tl{\hbox{\tiny L}\,}

\begin{document}
\title{Non-divisibility vs backflow of information in understanding 
revivals of quantum correlations for continuous-variable systems 
interacting with fluctuating environments}
\author{Jacopo Trapani}\email{jacopo.trapani@unimi.it}
\affiliation{Quantum Technology Lab, Dipartimento di Fisica, 
Universit\`a degli Studi di Milano, I-20133 Milano, Italy}
\author{Matteo G.~A.~Paris}\email{matteo.paris@fisica.unimi.it}
\affiliation{Quantum Technology Lab, Dipartimento di Fisica, 
Universit\`a degli Studi di Milano, I-20133 Milano, Italy}
\affiliation{CNISM, UdR Milano Statale, I-20133 Milano, Italy}
\affiliation{INFN, Sezione di Milano, I-20133 Milano, Italy}
\date{\today}
\begin{abstract}
We address the dynamics of quantum correlations for a bipartite 
continuous-variable quantum system interacting with its fluctuating 
environment. In particular, we consider two independent quantum 
oscillators initially prepared in a Gaussian state, e.g. a squeezed 
thermal state, and compare the dynamics resulting from {\em local
noise}, i.e. oscillators coupled to two independent external fields, 
to that originating from {\em common noise}, i.e. oscillators interacting 
with a single common field. We prove non-Markovianity (non-divisibility) 
of the dynamics in both regimes and analyze the connections between 
non-divisibility, backflow of information and revivals of quantum 
correlations. Our main results may be summarized as follows: (i) revivals 
of quantumness are present in both scenarios, however, the interaction 
with a common environment better preserves the quantum features of 
the system; (ii) the dynamics is always non-divisible but revivals of 
quantum correlations are present only when backflow of information is
present as well. We conclude that non-divisibility in 
its own is not a resource to preserve quantum correlations in 
our system, i.e. it
is not sufficient to observe recoherence phenomena. 
Rather, it represents a necessary prerequisite to
obtain backflow of information, which is the true ingredient to obtain
revivals of quantumness.
\end{abstract}
\maketitle
\section{Introduction}\label{s:intro}
Decoherence is a distinctive sign of the detrimental influence of 
the environment on a quantum system. In quantum information and
technology, decoherence is the main obstacle to reliable quantum 
processing of information. More generally, decoherence is a widely 
accepted explanation for the loss of nonclassicality
of quantum systems and for their transition to the classical
realm \cite{zur01, paz01}. In the recent years, it has been recognized 
that the action of an environment on a system may also have some 
non-detrimental effects, at least for a transient. Indeed, non-trivial 
spectral structures and memory effects, usually leading to 
non-Markovian dynamics for the quantum system 
\cite{liu02, smi01, pii04, smi02}, may induce recoherence and revivals 
of quantum features.
\par
The environment of a quantum system is usually made of several 
(classical and quantum) units with an overall complex structure. 
As a consequence, a full quantum treatment of the interaction betweeen 
a system and its environment is often challenging, or even unfeasible 
in a closed form. On the other hand, it is often the case that the 
overall action of the environment may be conveniently described as 
an external random force acting on the system, i.e. a classical 
stochastic field (CSF) \cite{gar01}.  
In general, one might expect that the modeling of a 
quantum environment as classical one leads to an incomplete 
description, i.e. such a description cannot capture the full 
quantum features of the dynamics. On the contrary, it 
has been proved that several system-environment interactions
have a classical equivalent description \cite{hel01, hel02, cro01, wit01,
str02, sto01}. In addition, there are experimental situations in which
quantum systems interact with an inherently classical Gaussian 
noise \cite{ast01, gal01, abe01, str01}, or where 
the environment can be effectively simulated classically
\cite{tur01}. 
\par
In this framework, the first goal of this paper is to analyze in details
the dynamics of a bipartite system made of two independent quantum
harmonic oscillators interacting with its classical fluctuating
environments. In particular, we compare the dynamics of correlations in
two different environmental situations. On the one hand, we consider a
local noise model, where each oscillator interacts with its own
classical environment. On the other hand, we also consider the situation
where both the oscillators interact with a common environment, described
by a single stochastic field.  A similar analysis has been performed for
qubit systems \cite{bel02, maz02} revealing a rich phenomenology.
\par
Our work is also aimed at better analyzing the connections between 
the dynamics of quantumness, e.g. revivals of quantum correlations 
and the quantum-to-classical transition, and the non-Markovian 
features of the dynamics. In particular, we want to investigate the 
role of non-Markovianity itself (i.e. non-divisibility of the 
quantum dynamical map) against the role of the backflow of information, 
which is a sufficient (but not necessary) condition to prove 
non-Markovianity and, in turn, often used to witness its presence.
\par
In order to introduce the subject, we remind that directly proving \cite{rhp, ill01, ade01}
the non-Markovian character of a dynamics is not always possible, 
as in many situations the full analytic form of the time-dependent
quantum dynamical map is missing. When the direct verification is 
not possible, one may exploit witnesses of non-Markovianity,  
i.e. quantities that vanish in case of a Markovian dynamics. Even though 
the witnesses may successfully capture the memory feature of a 
non-Markovian process in many situations, they possess 
different physical meaning and may be ineffective in some specific 
situation.  As an example, the BLP measure \cite{pii01}, and its 
continuous variable analogue based on fidelity \cite{nmc11}, have 
both a clear physical interpretation in terms of
information flow from the environment back to the system. 
Used as a witness, the BLP measure has been proved useful 
and effective to quantify 
memory effects in several situations but it may also fail to detect 
non-Markovianity \cite{hai01,pii02}. On the other hand, recent results  
\cite{hp1} suggest that information backflow is the essential element 
in addressing  non-Markovian dynamics of a quantum system and for this 
reason the BLP measure has been proposed 
as the definition of quantum non-Markovianity itself.
\par
In this context, more information may be often extracted from the study
of the quantum correlations of the system, e.g.  the entanglement or the
quantum discord. In fact, revivals of quantum features, especially in
non-interacting bipartite systems, may be a signature of a
non-Markovianity. This is what happens, e.g., for non-interacting qubits
\cite{bel01, bel02}. On the other hand, the connection between revivals
of correlations and backflow of information appears quite natural,
especially in open quantum systems: a temporary and partial restoration
of quantum coherence, previously lost during the interaction, is a sign
of a memory effect, possibly of the environment, which is supposed to be
storing correlations and sending them back to the system.  Remarkably,
revivals of correlations may also be found when the quantum system is
interacting with a classical environment, suggesting that these feature
reflects a property of the map, rather than a property of the
environment. For qubit systems revivals have been detected in presence
of Gaussian noise \cite{ben01}, and may even be found in the case of
non-interacting qubits \cite{rlf01}. For a single oscillator, classical 
memory effects have been found to increase the survival time of quantum 
coherence \cite{catstf} and, in particular, it has been proved that 
a detuning between the 
natural frequency of the system and the central frequency of the 
classical field induces revivals of quantum
coherence.
\par
For the system under investigation in this paper, non-Markovianity needs
not to be witnessed, as it can be easily proven in a direct way.  The
stochastic description of the environment, in fact, allows us to
determine the analytic form of the quantum dynamical map and to check
straightforwardly its non-divisibility. On the other hand, information
backflow may be revealed using fidelity witness \cite{nmc11}.  Our
results indicate that the dynamics is not divisible for both local and
common noise, and that also revivals of quantumness appears in both
cases, with the common case
better preserving the quantum features of the system. At the same time,
we found that revivals of quantum correlations are present only when
non-Markovianity is present {\em together} with a backflow of
information. We conclude that non-Markovianity itself is not the
resource needed to preserve quantum correlations in our system. In other
words, non-Markovianity alone is not a sufficient condition to induce
revivals.  Rather, it represents a necessary prerequisite to obtain
backflow of information, which is the true ingredients to obtain
revivals of quantumness.
\par
The paper is organized as follows. In Sec.\ref{s:inter} we introduce the
system and the stochastic modeling of both the environmental scenarios.
In Sec. \ref{s:correlations} we introduce the quantifiers of
correlations and describe the initial preparation of the system,
focusing on its inital correlations. In Sec. \ref{s:dynamics} we analyze
the dynamics of the correlations, delimiting the boundaries for the
existence of revivals. In Sec. \ref{s:nonmark} we prove the
nonmarkovianity of the dynamics and study the fidelity witness. Sec.
\ref{s:conclusions} summarizes the paper.
\section{The Interaction Model}
\label{s:inter}
We consider two non-interacting harmonic quantum oscillators with 
natural frequencies $\omega_1$ and $\omega_2$ and analyze the dynamics 
of this system in two different regimes: in the first one each 
oscillator is coupled to one of two independent non-interacting 
stochastic fields, we dub this scenario as the {\em local noise} case. 
In the second regime, the oscillators are coupled to the same classical 
stochastic field, so we dub this case as {\em common noise}.
In both case, the Hamiltonian $H$ is composed by a free and 
an interaction term. 
The free Hamiltonian $H_0$ is given by
\begin{align}
H_0 &= \hbar \sum_{j=1}^2  \omega_j  a_j \daga a_j 
\end{align}
Unlike the free Hamiltonian $H_0$, 
which is the same in the description of both models,
the interaction term $H_I$ differs. In the following subsections, we introduce
the local and the common interaction Hamiltonian.
\subsection{Local Interaction}
The interaction Hamiltonian $H_{\tl}$ in the local model reads
\begin{align}
H_{\tl} (t)& = \sum_{j=1}^2 a_j
\bar{C_j}(t) e^{i \delta_j t} + a\daga_j C_j (t) e^{-i \delta_j t}
\end{align}
where the annihilation operators $a_1,a_2$ represent the oscillators,
each coupled to a different local stochastic field $C_j (t)$ with $j =
1,2$, and $\delta_j = \omega_j-\omega$ is the detuning between the
carrier frequency of the field and the natural frequency of the $j$-th
oscillator. In the last equation and throughout the paper, we will
consider the  Hamiltonian rescaled in units of a reference level of
energy $\hbar \omega_0$ (for a reason to be pointed out later).  Under
this condition, the stochastic fields $C_1 (t), C_2(t)$, their central
frequency $\omega$, the interaction time $t$, and the detunings all 
become dimensionless quantities.
\par
The presence of fluctuating stochastic fields leads to 
an explicitly time-dependent Hamiltonian, whose corresponding 
evolution operator is given by
\begin{align}
\label{eq:evolutionoperator}
U(t) = {\cal T} \exp\left\{ - i \int_0^t \!\! ds\, H_{\tl}(s) \right\}\,,
\end{align}
where ${\cal T}$ is the time ordering.
However, as the interaction Hamiltonian is linear
in the annihiliation and creation operators of the two oscillators,
 the two-time commutator $[H_{\tl}(t_1),H_{\tl}(t_2)]$ is always proportional 
to the identity. 
In particular, when the stochastic fields satisfy the conditions
$C_j(t_1) \bar{C_j}(t_2) = C_j(t_2) \bar{C_j}(t_1)$, with $j = 1,2$,
the two-time commutator becomes
\begin{align}
[H_{\tl} (t_1), H_{\tl} (t_2)]  = & \sum_{j=1,2} 2 i \, \bar{C_j}(t_1) C_j(t_2)
\notag \\ & \times \sin\left[ \delta_j(t_1 -t_2)\right]\, \mathbb{I}_{12}
\end{align}
This form of the two-time commutator allows us 
to use the Magnus expansion \cite{mag1,mag2} to simplify
the expression of the evolution operator (\ref{eq:evolutionoperator}) into
\begin{align}
\label{eq:evolutionoperator2}
U(t) = \exp \left( \Xi_1 + \Xi_2 \right)
\end{align}
where the most general form of $\Xi_1$ and $\Xi_2$ is given by 
\begin{align}
\Xi_1 &= -i \int_0^t\!\! ds_1\, H_I (s_1), \\
\Xi_2 &=  \frac12 \int_0^t\!ds_1\!\int_0^{s_1}\!\! ds_2\,
\, [H_I(s_1),H_I(s_2)].
\end{align}
The specific form of $\Xi_1$ for the local ($\Xi_1^{\tl}$) scenario is given by
\begin{align}
\Xi_1^{\tl} &= \sum_{j=1}^2 \left(a_j\daga \phi_j (t)  - a_j \phi_j^{*} (t)\right)
\end{align}
where 
\begin{align}
\phi_j (t) &= -i \int_0^t ds \, e^{-i \delta_j s} C_j (s) \quad \mbox{with } j=1,2.
\end{align}
The evolution of the density operator of the system then reads
\begin{align}
\rho_L (t) &= \left[e^{\Xi^{\tl}_1} \rho(0) e^{(\Xi_1^{\tl})^{*}}
\right]_{F} \notag \\ &=  
\left[D(\phi_a, \phi_b)\rho_0 D\daga (\phi_a, \phi_b) \right]_{F}  
\end{align}
where $D_j(\alpha) = \exp(\alpha a_j\daga - \alpha^{*}a_j)$ is the displacement operator, $D(\alpha_1, \alpha_2) = D(\boldsymbol{\alpha}) = D_1(\alpha_1) D_2 (\alpha_2)$ and $\left[\ldots \right]_{F}$ is the average over the realizations of the stochastic fields.
\par
In the local scenario, we assume each CSF 
$$C_j(t) = C_j^{(x)}(t) + i C_j^{(y)}(t),$$
described as a Gaussian stochastic process
with zero mean 
$ [C_j^{(x)}(t)]_{F} =  [C_j^{(y)}(t)]_{F}=0$ and 
autocorrelation matrix given by
\begin{align}
\left[C_j^{(x)}(t_1)C_k^{(x)}(t_2) \right]_F &= 
\left[C_j^{(y)}(t_1)C_k^{(y)}(t_2) \right]_F \notag \\ & = \delta_{jk} K(t_1,t_2) \\
\left[C_j^{(x)}(t_1)C_k^{(y)}(t_2) \right]_F 
&= \left[C_j^{(y)}(t_1)C_k^{(x)}(t_2) \right]_F =0 
\end{align}
where we introduced the kernel autocorrelation function $K(t_1,t_2)$. 
By means of the Glauber decomposition of the initial state $\rho(0)$
\be
\rho(0) = \int \frac{d^4 \boldsymbol{\zeta}}{\pi^2} \chi [\rho(0)](\boldsymbol{\zeta}) D\daga(\boldsymbol{\zeta}) 
\ee
where $\chi [\rho] (\boldsymbol{\zeta}) = \mbox{Tr}[\rho D(\boldsymbol{\zeta})]$ is the 
symmetrically ordered characteristic function, the density matrix of the evolved
state reads
\be
\label{locdyn}
\rho_{\tl}(t) = \mathcal{G}_{\tl}[\rho(0)] = \int \frac{d^4 \boldsymbol{\zeta}}{\pi^2}g_{\tl}(\boldsymbol{\zeta}) D(\boldsymbol{\zeta})\rho(0) D\daga(\boldsymbol{\zeta})
\ee
where we use the Gaussian function
\be
g_{\tl}(\boldsymbol{\zeta})= \frac{\exp(-\frac12 \, \boldsymbol{\zeta} \cdot \boldsymbol{\Omega} \cdot \boldsymbol{\sigma}_{\tl}^{-1}\cdot \boldsymbol{\Omega}^T \cdot \boldsymbol{\zeta}^T)}{\sqrt{\mbox{det}[ \boldsymbol{\sigma}_{\tl}]}}
\ee
where $\boldsymbol{\sigma}_{\tl}$ 
and the symplectic matrix $\boldsymbol{\Omega}$ are given by
\be 
\boldsymbol{\Omega} = \left(
\begin{array}{cc}
0 & 1 \vspace{0.1 cm}
\\

-1 & 0 
\end{array}
\right)
\quad
\boldsymbol{\sigma}_{\tl} = \left(
\begin{array}{cc}
\beta_1 (t) \mathbb{I}_2 & 0 \vspace{0.1 cm} \\
0 & \beta_2(t) \mathbb{I}_2
\end{array} \right).
\label{sigl}
\ee
The matrix $\boldsymbol{\sigma}_{\tl}$  is the covariance of the noise 
function $g_{\tl}(\boldsymbol{\sigma})$
and its matrix elements are given by 
\be
\beta_j (t,t_0)= \int_{t_0}^t \int_{t_0}^t d s_1 d s_2 
\cos[\delta_j(s_1 - s_2)] K(s_1,s_2).
\ee
\subsection{Common Interaction}
The Hamiltonian $H_{\tc}$ in the common interaction model reads
\begin{align}
H_{\tc} (t)& = \sum_{j=1}^2 a_j  e^{i \delta_j t}\bar{C}(t) + a\daga_j  e^{-i \delta_j t}C (t)
\end{align}
where each oscillator, represented by the annihilation operators $a_1,a_2$, is coupled to a common stochastic
field $C(t)$ which is described as a Gaussian stochastic process with zero mean $[C^{(x)}]_F = [C^{(y)}]_F =0 $ and  the very same
autocorrelation matrix of the local scenario.
\par
Along the same lines of the local interaction model derivation, we use the Magnus expansion in order to get to the evolution operator. 
By asking the stochastic field to satisfy the relation $C(t_1) \bar C(t_2) = C (t_2) \bar C(t_1)$, the two-time commutator reads
\begin{align}
[H_{\tc} (t_1), H_{\tc} (t_2)] &= \bar{C}(t_1) C(t_2)  \sum_{j=1,2} 2 i \, \sin\left[ \delta_j(t_1 -t_2)\right] \mathbb{I}_{12}.
\end{align}
The evolution operator for the common scenario is the same described in
Eq. (\ref{eq:evolutionoperator2}), where the specific form of $\Xi_1$ in the common interaction model is given by
\begin{align}
\Xi_1^{\,\tc} &= \sum_{j=1}^2 \left(a_j\daga \psi_j (t)  - a_j \psi_j^{*} (t)\right)
\end{align}
where
\begin{align}
\psi_j (t) & = -i \int_0^t ds \, e^{-i \delta_j s} C (s) \quad \mbox{with } j=1,2. 
\end{align}
The evolution of the density operator of the system then reads
\begin{align}
\rho (t) = \left[e^{\Xi^{\, \tc}_1} \rho(0) e^{(\Xi^{\, \tc}_1)^{*}} \right]_{F} = 
\left[ D(\psi_1, \psi_2)\rho_0 D\daga (\psi_1, \psi_2) \right]_{F}   
\end{align}
which, following the same steps of the derivation presented in the previous subsection, leads to
\be
\label{comdyn}
\rho_{\tc}(t) = \mathcal{G}_{\tc}[\rho(0)] = \int \frac{d^4 \boldsymbol{\zeta}}{\pi^2}g_{\tc}(\boldsymbol{\zeta}) D(\boldsymbol{\zeta})\rho(0) D\daga(\boldsymbol{\zeta})
\ee
where we use the Gaussian function
\be
g_{\tc}(\boldsymbol{\zeta})= \frac{\exp(-\frac12 \, \boldsymbol{\zeta} \cdot  \boldsymbol{\Omega} \cdot \boldsymbol{\sigma}_{\tc}^{-1}\cdot \boldsymbol{\Omega}^T \cdot \boldsymbol{\zeta}^T)}{\sqrt{\mbox{det}[ \boldsymbol{\sigma}_{\tc}]}}
\ee
 $\boldsymbol{\sigma}_{\tc}$ being its covariance matrix, 
given by
\begin{align}
&\boldsymbol{\sigma}_{\tc} = \left(
\begin{array}{cc}
\beta_1 (t) \mathbb{I}_2 & \boldsymbol R \vspace{0.1 cm} \\
\boldsymbol R & \beta_2(t) \mathbb{I}_2
\end{array} \right)
\label{sigc}
\\
&\boldsymbol R = \left(
\begin{array}{cc}
\beta_{\tc} (t)  & \gamma_{\tc} (t) \vspace{0.1 cm} \\
\gamma_{\tc} (t) & \beta_{\tc}(t) 
\end{array} \right)
\end{align}
with the matrix elements given by
\begin{flalign}
\beta_c(t,t_0)&= \int_{t_0}^t \int_{t_0}^t d s_1 d s_2 \cos[(\delta_1 s_1 - \delta_2 s_2)] K(s_1,s_2), \nonumber
\\
\gamma_c(t,t_0)&= \int_{t_0}^t \int_{t_0}^t d s_1 d s_2 \sin[(\delta_1 s_1 - \delta_2 s_2)] K(s_1,s_2).
\end{flalign}
\subsection{Covariance Matrix dynamics in local and common interaction}
The dynamical maps described by Eqs. (\ref{locdyn}) and (\ref{comdyn}) 
correspond to Gaussian channels, i.e. the evolution, 
in both regimes, preserves the Gaussian character of the input state. 
In turn, this is a useful features, since in this case quantum 
correlations, entanglement and discord, may be evaluated exactly.
We also remind that Gaussian channels represent the short times
solution of Markovian (dissipative) Master equations in the limit of
high-temperature environment. In the following, this link will be
exploited to analyze the limiting behaviour of the two-mode dynamics.
\par
In order to get quantitative results, we assume that fluctuations 
in the environment are described by Ornstein-Uhlenbeck Gaussian 
processes, characterized by a Lorentzian spectrum and
a kernel autocorrelation function 
$$K(t_1,t_2) = \frac{1}{2}\lambda t_{\es}^{-1} \exp({-|t_1 -t_2|/t_{\es}})\,,$$
where $\lambda$ is a coupling constant and $t_{\es}$ is the
 correlation time of the environment. 
We also assume the case of resonant oscillators
$(\omega_1 = \omega_2 = \omega_0)$, which implies that 
the oscillators are identically detuned
from the central frequency of the classical stochastic field, 
i.e. $$\delta_1 = \delta_2 = \delta = 1-\frac{\omega}{\omega_0}\,.$$
This assumptions simplifies the expression of the state dynamics:
in the local scenario, leading to
$\beta_1 (t) = \beta_2 (t) = \beta(t)$ and, in turn, 
\begin{flalign}
\label{eq:symplocdyn}
\rho_{\tl} (t) &= \mathcal{E}_{\tl}[\rho(0)](t) =\nonumber \\
&=\int \frac{d^4 \boldsymbol \zeta}{(\pi \beta(t))^2} 
\exp \left(- \frac{|\boldsymbol \zeta|^2}{\beta(t)} 
\right)  D(\boldsymbol \zeta) \rho(0) D\daga (\boldsymbol \zeta)
\end{flalign}
where $\beta(\Delta t = t- t_0) = \beta(t,t_0)$ with the 
Ornstein-Uhlenbeck kernel is
\begin{align}
\label{beta}
\beta(t) =& \frac{\lambda}{[ 1 + (\delta\, t_{\es})^2]^2}
\Big\{t- t_{\es} + (\delta\, t_{\es})^2 (t+ t_{\es})\notag \\ &+ t_{\es}\, 
e^{-t/t_{\es}}\left[(1-(\delta\, t_{\es})^2)\cos \delta\, t - 
2 \delta\, t_{\es} \sin \delta\, t\right]\!\Big\}.
\end{align}
In the common noise case, the condition of resonant oscillators 
implies $\beta_1(t) = \beta_2(t) = \beta_c(t) = \beta(t)$ and
$\gamma_c (t) =0$, leading to simplified matrices 
$\boldsymbol R$ and $\boldsymbol \sigma_{\tc}$ given by
\be
\boldsymbol R = \left(
\begin{array}{cc}
\beta (t)  & 0\vspace{0.1 cm} \\
0 & \beta (t) 
\end{array} \right)
\quad
\boldsymbol{\sigma}_{\tc} = \left(
\begin{array}{cc}
\beta (t) \mathbb{I}_2 & \boldsymbol R \vspace{0.1 cm} \\
\boldsymbol R & \beta(t) \mathbb{I}_2
\end{array} \right)
\label{sigc1}
\ee
corresponding to the Gaussian channel
\begin{align}
\label{eq:sympcomdyn}
\rho(t) &= \mathcal{E}_{\tc}[\rho(0)](t) =   \nonumber \\ &=  \int \frac {d^2 \zeta }{\pi \beta(t) } \exp \left( - \frac{|\zeta|^2}{\beta(t)}\right) D(\zeta,\zeta) \rho(0) D\daga(\zeta,\zeta).
\end{align}
As for the initial state $\rho(0)$ of the system, we assume a generic 
squeezed thermal state (STS), which is a zero-mean Gaussian state, 
described by a Gaussian characteristic function 
$\chi[\rho(0)](\boldsymbol \zeta) = \exp \left( -\frac12 \boldsymbol
\zeta^T  \boldsymbol \sigma_{\sts} \boldsymbol \zeta   \right) $
where the covariance matrix $\boldsymbol \sigma_{\sts}$ of the state has the general form
\be
\label{sqthermalstate}
\boldsymbol \sigma_{\sts} =\frac12 \left(
\begin{array}{cc}
A\, \mathbb I_2 &  C \sigma_z \\
C \sigma_z  &  B\, \mathbb I_2 
\end{array} \right),
\ee
where $\sigma_z = \mbox{diag}(1,-1)$ is the $z$ Pauli matrix.
This covariance matrix corresponds to a density operator of the form
\be
\rho = S_2 (r) (\nu_1 \otimes \nu_2) S_2 (r)\daga
\ee 
where $S_2 (r) = \exp \{ r (a_1\daga a_2\daga -a_1 a_2) \}$ is the two-mode squeezing operator and $\nu_j$
is a single-mode thermal state 
\be
\nu_j = \frac{1}{\bar n_j}\sum_m \left(\frac{\bar n_j}{\bar n_j +1}\right)^m |m \rangle \langle m|.
\ee
The physical state depends on three real parameters: 
the squeezing parameter $r$ and the two numbers $\bar n_1,\bar n_2$, 
which are related to the parameters $A,B,C$ of eq. (\ref{sqthermalstate}) by the relations
\begin{align} 
A &= \cosh (2r) + 2 \bar n_1 \cosh^2 r + 2 \bar n_2 \sinh^2 r \nonumber \\
B &= \cosh (2r) + 2 \bar n_1 \sinh^2 r + 2 \bar n_2 \cosh^2 r \nonumber \\
C &= (1+ \bar n_1 + \bar n_2 ) \sinh (2 r).
\end{align}
\par
As the squeezed thermal state is a Gaussian state and the dynamics in
both scenarios is described by a Gaussian channel, the output state at
any time is Gaussian as well, so it is determined only by the covariance
matrix. By evaluating the characteristic function of the evolved state,
one finds that the covariance matrices of the state at time $t$ in the
local and common scenarios are 
\begin{align}
\bbeta_{\tl} (t) &= \bsigma_{\sts} + 2 \, \bsigma_{\tl} (t) \label{evsl} \\
\bbeta_{\tc} (t) &= \bsigma_{\sts} + 2 \, \bsigma_{\tc} (t) \label{evsc}
\,,
\end{align}
where
$\bsigma_{\tl} (t)$ and
$\bsigma_{\tc} (t)$ are given in Eqs. (\ref{sigl}) and (\ref{sigc1})
respectively.
As mentioned before, the Gaussian character of the states is preserved
by the dynamics. Nevertheless, while the output state in the local
scenario is always a STS, i.e. at any time it can be put in a
diagonal form by means of a two-mode squeezing operation, the output
state in the common scenario ceases to be an STS as soon as the 
interaction starts.
\section{Quantum Correlations}
\label{s:correlations}
Systems possessing quantum features have proved themselves useful in
many fields, e.g. quantum information, quantum computation and quantum
estimation, increasing the performances of computation protocols and
precision measurements. In the case of bipartite systems, the
quantumness lies both in the nonclassicality of the state and in the
correlations between the different parts of the system. In this section,
we introduce two well-known markers of correlations, entanglement and
quantum discord.  
\subsection{Entanglement}
While classical multipartite systems, though being correlated, are
always described in a quantum picture by separable states, quantum
multipartite systems may show nonclassical correlations which require a
description in terms of non-separable density operators. Aiming at
quantifying these quantum correlations, the entanglement measures the
degree of non-separability of a quantum system. For a bipartite Gaussian
state $\rho$ with covariance matrix $\sigma$, the entanglement is given
by the logarithmic negativity, which is defined as \be
\label{eq:entanglement}
\mathcal N (\rho) = \mbox{max} \{0, -\log (2 \tilde d_{-})\}
\ee
where $\tilde d_{-}$ is the smallest symplectic eigenvalue of the partially transposed covariance matrix $\sigma^{\hbox{\tiny PT}\,} = \Delta \sigma \Delta$ with $\Delta = \mbox{diag}(1, -1,1,1)$. As stated by Simon \cite{sim1}, a two-mode gaussian state is separable if and only if the symplectic eigenvalue satisfies the relation $\tilde d_{-}\geq \frac12$. Any violation of the latter implies that the state is non-separable (or entangled) and leads to a positive measure of the entanglement given in (\ref{eq:entanglement}).
\begin{figure}[h!]
\centering
\includegraphics[width=0.48 \columnwidth]{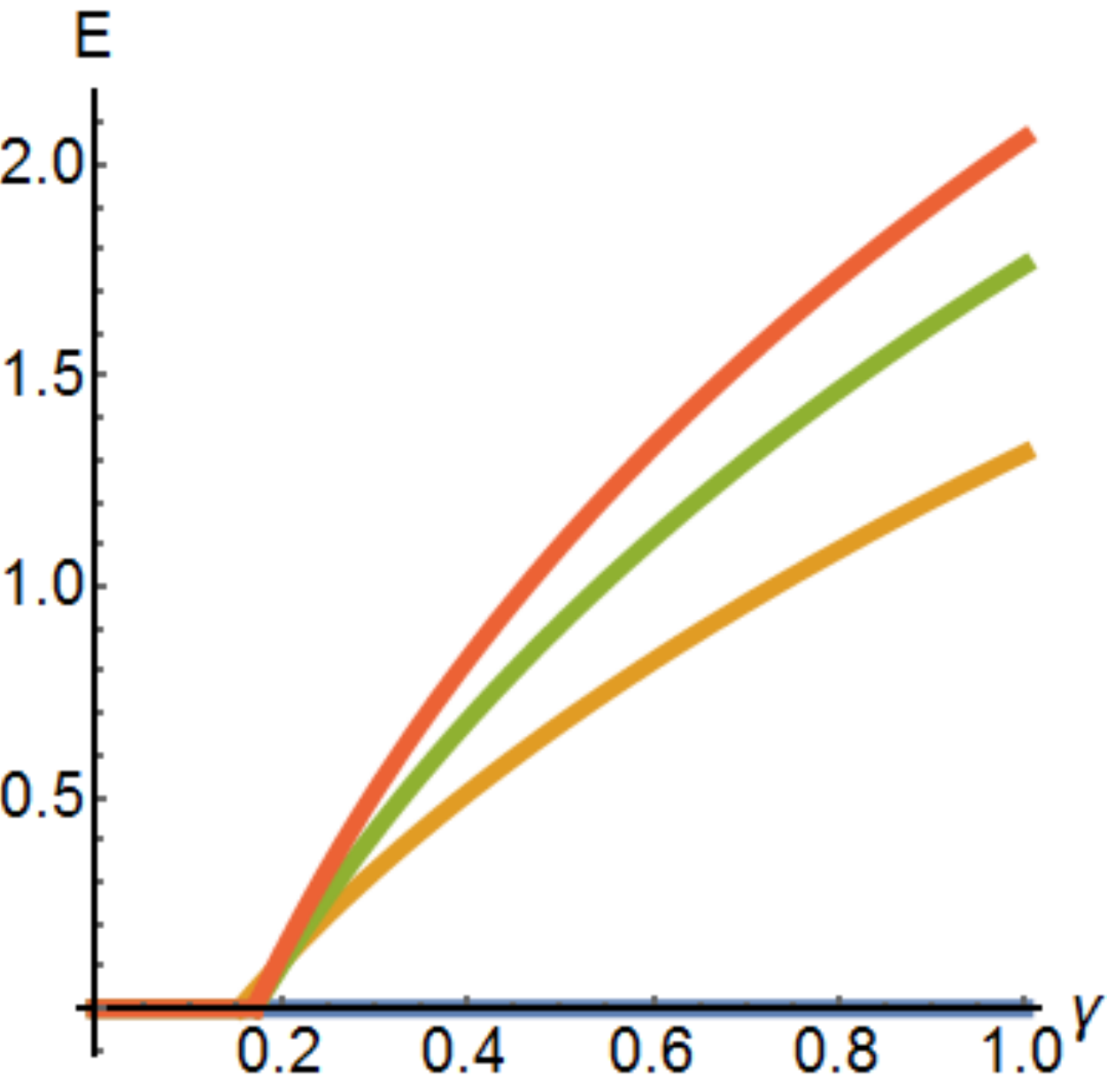}
\includegraphics[width=0.48 \columnwidth]{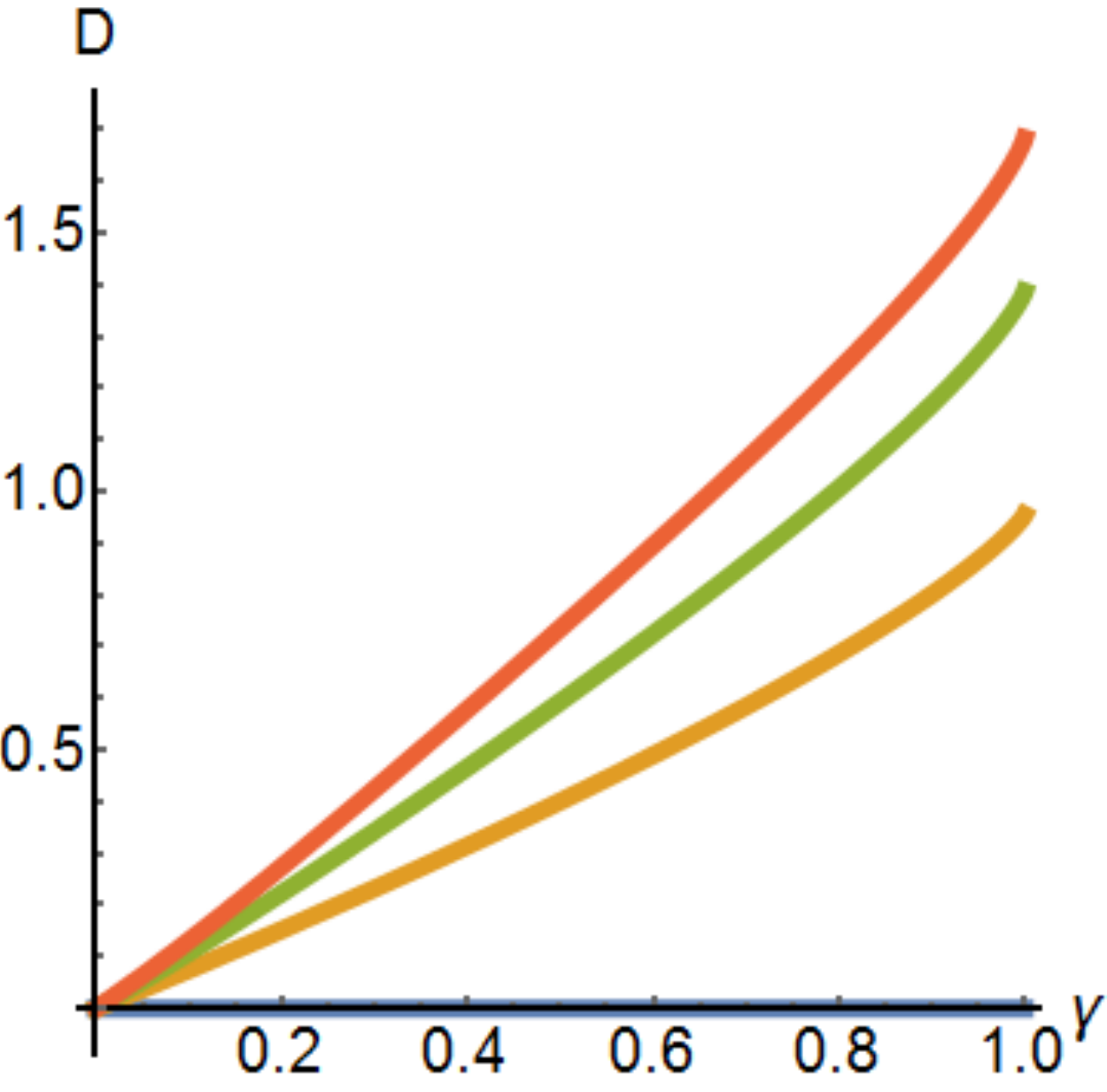}
\caption{Quantum correlations of STS for different values of 
the energy $\epsilon$. Left panel: Entanglement of a STS as a 
function of squeezing parameter $\gamma$. The STS is entangled 
as long as $\gamma$ overtakes a threshold value that depends on 
the energy $\epsilon$. Right panel: Discord of a STS as a function 
of squeezing parameter $\gamma$. The STS is always a discordant state 
unless $\gamma = 0$. In both panels, from bottom to top, 
$\epsilon =0$ (blue line), , $\epsilon =1$ (yellow line) , 
$\epsilon =2$ (green line) , $\epsilon =3$ (red line)  }\label{f:STS1}
\end{figure}
\subsection{Quantum Discord}
The total amount of the correlations possessed by a bipartite quantum state $\rho$ is called mutual information and is given by
\be
\mathcal{I(\rho)} = S(\rho_1) + S(\rho_2) - S(\rho),
\ee
where $S(\rho_j)$ is the Von Neumann entropy of the $j$-th subsystem. Usually, the mutual information can be divided into two parts: a classical part $\mathcal C (\rho)$ and a quantum part $\mathcal D (\rho)$, which takes name of quantum discord.
The classical correlations, defined as the maximum amount of information extractable from one subsytem by performing local operations on the other, are given by
\be
\mathcal C (\rho) = \mbox{max}_{\Pi_i} \Big\{ S(\rho_1) - \sum_i p_i S(\rho^{\Pi_i}_{1|2}) \Big\}
\ee
where $\rho^{\Pi_i}_{1|2} = \mbox{Tr}_2 (\rho \mathbb I \otimes \Pi_i)$ is the state after the measurement on system 2 with
probability $p_i = \mbox{Tr}_{1,2} (\rho \mathbb I \otimes \Pi_i) $.
The quantum discord is defined as the difference between the total correlations and the classical correlations:
\be
\mathcal{D}(\rho) = \mathcal I (\rho) - \mathcal C (\rho).
\ee
The quantum discord then measures the amount of correlations whose
origin cannot be addressed to the action of local operations or
classical communication. However, computing the quantum discord may be
challenging as it usually implies finding the POVM that maximizes the
classical correlations. In the case of Gaussian states, the form of the
POVM maximizing the classical correlations is known \cite{gio0,gio1} and the
quantum discord depends only on the covariance matrix by the relation
\be
\label{eq:discord}
\mathcal D (\rho) = h(\sqrt{I_2}) - h(d_{-}) - h(d_{+})  + h\left( \sqrt{E_{min}} \right),
\ee
where $d_{-}$ and $d_{+}$ are the symplectic eigenvalues of the
covariance matrix, $I_1, I_2, I_3, I_4$ are the so-called symplectic
invariants, $h(x)= (x+\frac12) \log (x+ \frac12)
-(x-\frac12)\log(x-\frac12)$ and \cite{gio1} 
\be
E_{min} = \begin{cases} \left[ \frac{2 |I_3| +\sqrt{4 I_3^2+ (4 I_2
-1)(4 I_4 -1)}}{4 I_2 -1} \right]^2 & \mbox{if } R_\sigma \leq 1 \\ \frac{I_1
I_2 + I_4 -I_3^2 - \sqrt{(I_1 I_2 + I_4 -I_3^2)^2 -4 I_1 I_2 I_4}}{2
I_2} & \mbox{if } R_\sigma > 1  \end{cases} \ee
where 
$$R_\sigma= \frac{4(I_1 I_2 -I_4)^2}{(I_1 + 4 I_4)(1+4I_2) I_3^2}\,.$$ 
For Gaussian states satisfying the second condition, 
the maximum amount of extractable information is achieved by measuring 
a canonical variable (e.g. by homodyne detection in optical systems
\cite{hmqd}). On the other hand, for states falling in the first set, 
the optimal measurement is more general, and coincides with 
the projection over coherent states for STSs. For
a generic Gaussian state, with covariance matrix $\sigma$ written in 
a block form
\be
\sigma =\left(
\begin{array}{cc}
\mathbb A & \mathbb C \\
\mathbb C^T & \mathbb B
\end{array} \right)
\ee
the symplectic invariants are  $I_1 = \mbox{det} \, \mathbb A$, $I_2 =
\mbox{det} \, \mathbb B$, $I_3 = \mbox{det} \, \mathbb C, I_4 =
\mbox{det}\, \sigma$.  
\subsection{STS correlations}
In order to assess the dynamics of entanglement and discord in the 
presence of noise, we briefly review the static properties of quantum 
correlations \cite{bru15} for a squeezed thermal state.
We consider the case of identical thermal states $(\bar n_1 = \bar n_2 =
\bar n)$ and use a convenient representation of STSs, built upon 
re-parametrizing the covariance matrix by means of its total energy 
$\epsilon = 2( \bar n + n_s +2 \bar n\, n_s
)$, with $n_s = \sinh^2 r$, and a normalized squeezing parameter $\gamma
\in [0,1]$, such that
\begin{align}
n_s &= \gamma \epsilon \nonumber 
\quad \quad \bar n = \frac{(1-\gamma) \epsilon}{1+2 \gamma \epsilon}.
\end{align}
Note that, for $\gamma = 0$ the state has only thermal energy ($\epsilon
= \bar n$) while for $\gamma = 1$ the total amount of energy comes from
the two-mode squeezing operation ($\epsilon = \sinh^2 r$).  \par
Fig.\ref{f:STS1} shows the quantum correlations of a STS as a function of the
energy $\epsilon$ and the squeezing parameter $\gamma$.  The left panel 
shows that the STS is entangled as long as $\gamma$ overtakes 
a threshold value which
depends on the total amount of energy. Conversely, the quantum discord
of a STS is always positive, unless the state is purely thermal, i.e. 
with zero squeezing.
($\gamma=0$).   
Notice that the states considered in this paper belongs to the class
of Gaussian states for which the Gaussian discord equals the full
quantum discord \cite{pir14}.
\section{Dynamics of Quantum Correlations}
\label{s:dynamics} 
Before proceeding with a detailed analysis of the dynamics of the quantum
correlations, let us focus on the function $\beta(t)$, in order to
understand which parameters affect the dynamics of the output
state.  As a matter of fact, the function $\beta(t)$, in (\ref{beta}) 
dependent on only two parameters
(except time $t$), as it can be rescaled in units of $t_{\es}$ by assuming
$\tilde{\delta} = \delta t_{\es}$, $\tilde{\lambda} = \lambda\, t_{\es}$,
$\tilde{t} = t/t_{\es}$, leading to the expression (in which tildes have
already been dropped) 
\begin{align}
\beta(t) =& \, \frac{\lambda}{[ 1 + \delta^2]^2}
\Big\{t-1 + \delta^2 (t+ 1)\notag \\ &+\, 
e^{-t}\left[(1-\delta^2)\cos \delta\, t - 
2 \delta\, \sin \delta\, t\right]\!\Big\}.
\end{align}
Intuitively, given the form of the covariance matrices in Eqs. 
(\ref{evsl}) and (\ref{evsc})
one realizes that when $\beta(t)$ shows a 
monotonous behaviour, the system cannot gain any quantum features, or 
go back to the initial 
state at any value of the interaction time. 
Conversely, an
oscillating $\beta(t)$ would let the system orbiting in the parameter
space, which means that the quantum features of the output state may have a
chance to be restored.  \par
\begin{figure}[h!]
\centering
\includegraphics[width=0.48 \columnwidth]{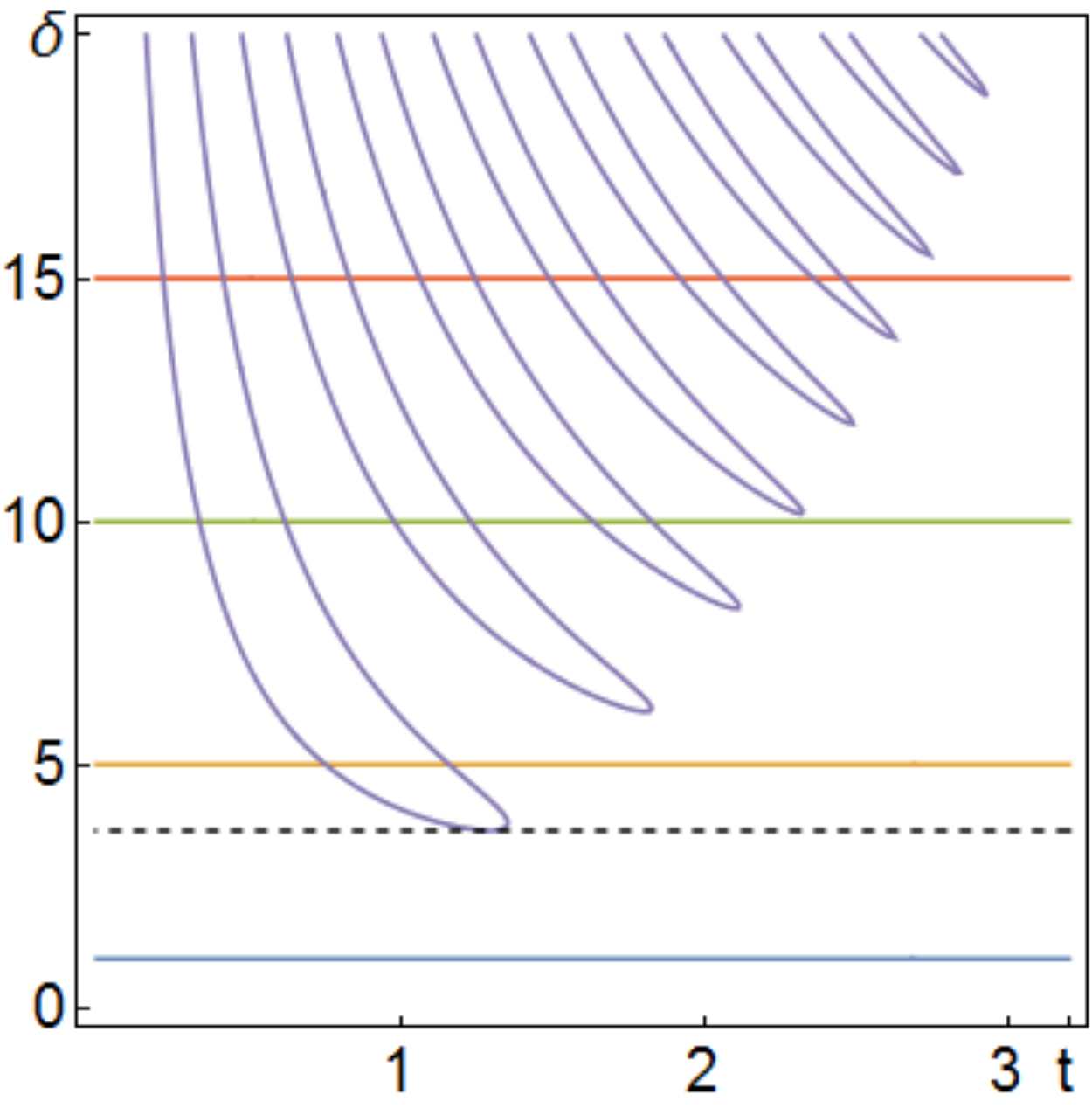}
\includegraphics[width=0.48 \columnwidth]{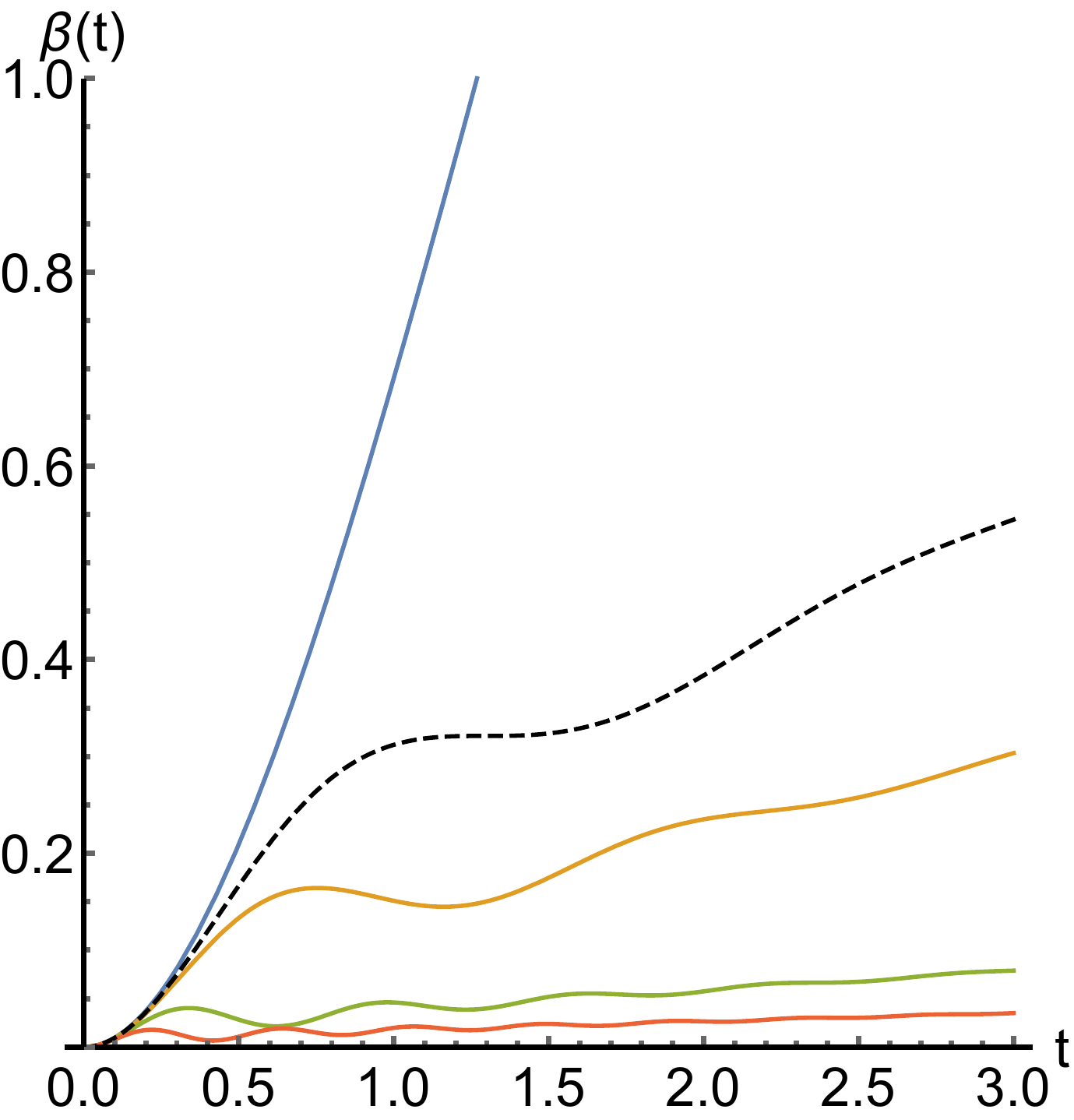}
\caption{Left panel: Contourplot of $d \beta(t) / dt = 0$ as a function of $t$ and $\delta$. The purple curve represents the solution of $d \beta(t) / dt = 0$. The black dashed line represents the maximum value of $\delta = \delta_0$ for which $\beta(t)$ does not oscillate. Right panel: $\beta(t)$ for different values of $\delta$. From bottom to top, the other lines are for $\delta =1$ (blue), $\delta = 5$ (yellow), $\delta =10$ (green), $\delta =15$ (red). An oscillating behavior is present only if $\delta>\delta_0$.  }\label{f:beta}
\end{figure}

Formally, upon imposing the condition $d \beta(t) / dt = 0$ leads to the
following equation \be
\label{eq:condition}
2 \lambda\, \frac{1-e^{-t}(\cos
\delta t - \delta \sin \delta t)}{1 + \delta^2} = 0
\ee
which can not be analitically solved. The left panel of Fig.
$\ref{f:beta}$ contains a numerical plot of the solutions of
(\ref{eq:condition}) and shows the existence of a lower bound on the
rescaled detuning $\delta$ for the oscillations of $\beta(t)$. The lower
bound is represented by the black dashed line, corresponding to
$$\delta_0 = \frac{3\pi}{2}[\mbox{\small{ProductLog}}(
\frac{3\pi}{2})]^{-1}  \simeq 3.644\,,$$ independently of $\lambda$.
\par
With this in mind, we now examinate the dynamics of quantum correlations
of initially maximally entangled squeezed thermal states ($\gamma =1$) and two-mode thermal states ($\gamma =0)$
 in presence of local and common stochastic environments. In order to be able of comparing the results of the different scenarios, we remind we already limited the analysis to resonant oscillators and that
we assume identical the rescaled coupling constant $\lambda^{(c)}$ for the common scenario and $\lambda^{(1)},\lambda^{(2)}$ for the local scenario, $\lambda^{(c)}= \lambda^{(1)} = \lambda^{(2)} = \lambda$.
\par
\begin{figure}[h!]
\centering 
\includegraphics[width=0.48 \columnwidth]{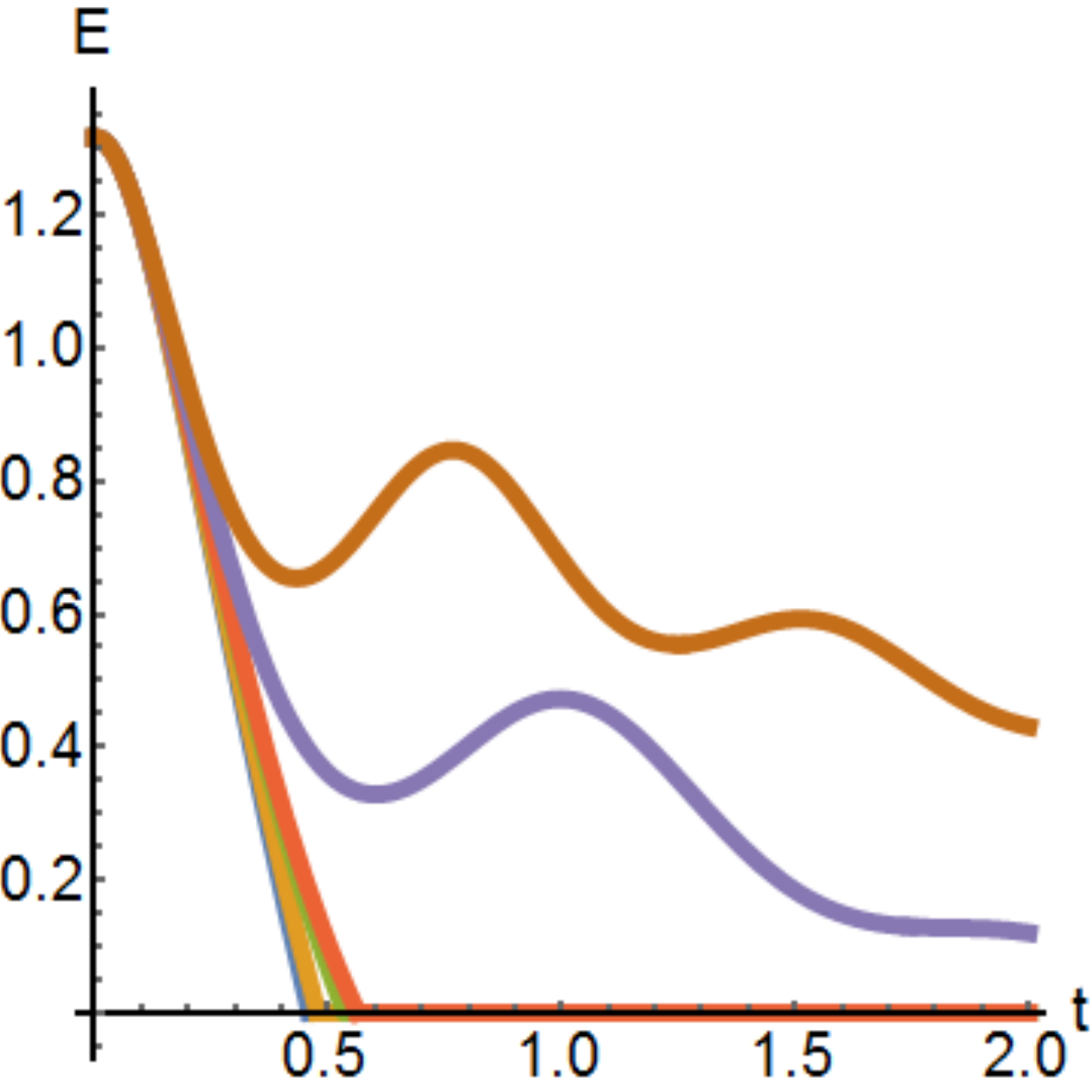}
\includegraphics[width=0.48 \columnwidth]{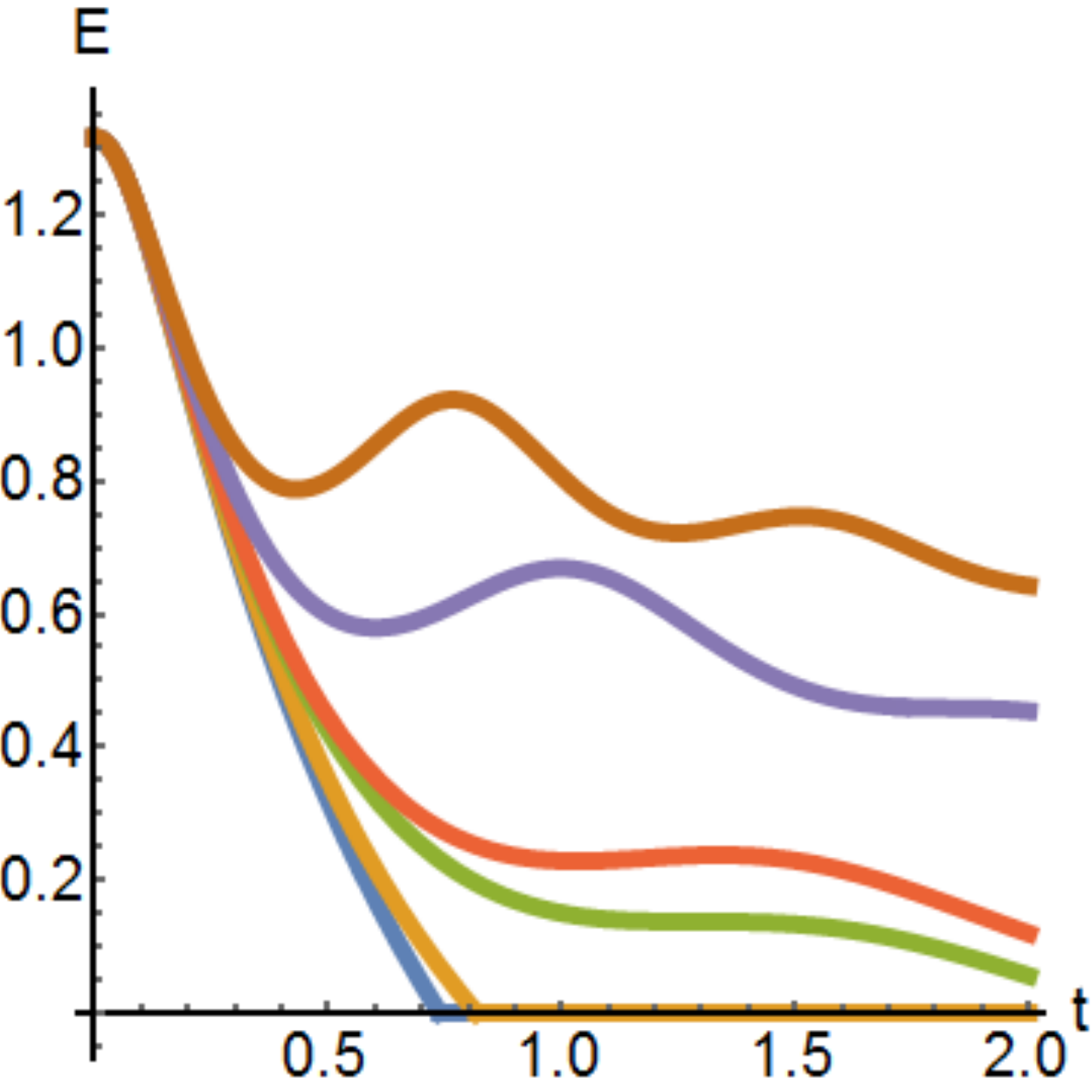}
\includegraphics[width=0.48 \columnwidth]{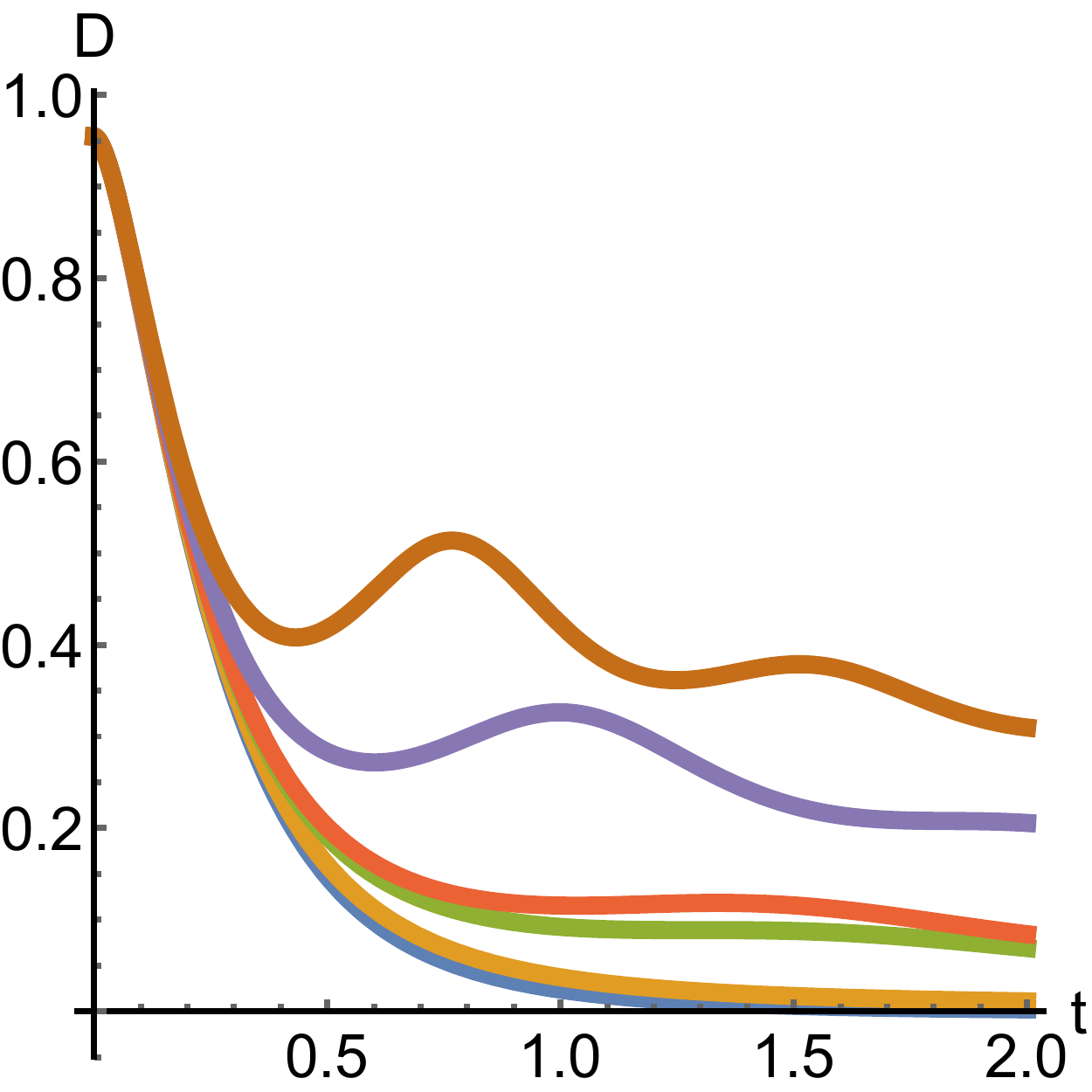}
\includegraphics[width=0.48 \columnwidth]{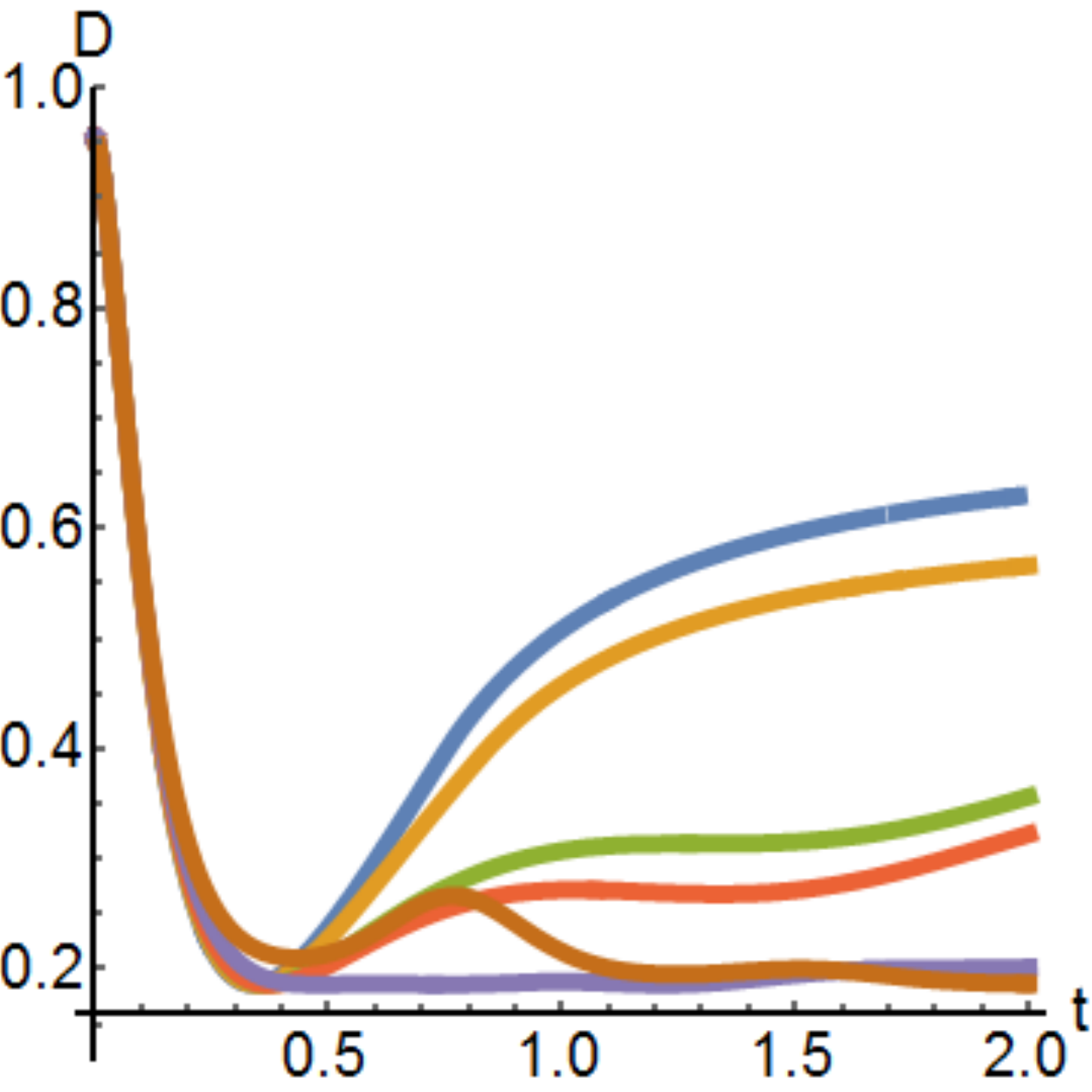}
\caption{(Color Online) Dynamics of correlations in presence of CSFs for
different values of rescaled detuning $\delta$.  Upper left panel:
Entanglement dynamics in local scenario as a function of time $t$. Upper
right panel: Entanglement dynamics in common scenario as a function of
time $t$. In both scenarios, the entanglement possessed by the system
may revive. Lower left panel: Discord dynamics in local scenario as a
function of time $t$. The initial discord decreases in time. Right left
panel: Discord dynamics in common scenario as a function of time $t$.
The initial discord decreases, reaches a minimum and the increases
monotonically as a consequence of the interaction.  In all panels, we
set $\epsilon =1, \gamma =1, \lambda=1$, and, from bottom to top,
$\delta = 0$ (blue line), $\delta = 2$ (yellow line), $\delta =
\delta_0$ (green line), $\delta = 4$ (red line), $\delta = 6$ (purple
line), $\delta = 8$ (brown line). }
\label{fig:ent1}
\end{figure}

Let us start by addressing 
the dynamics of correlations of an initially entangled
STS: the upper panels  in Fig. \ref{fig:ent1} show how the classical
stochastic fields, whether they be local or common, induce loss of
correlations in time. However, the decay rate of correlations is not the
same in both scenarios: indeed, the presence of a common stochastic
field is less detrimental, i.e. the interaction with the same
environment leads to a slower loss of correlations. In all the four panels,
the green line corresponds to $\delta = \delta_0$, the threshold value
over which $\beta(t)$ shows an oscillating behavior. As it is possible
to see, $\delta = \delta_0$ plays the role of the threshold value also
in the case of the correlations. In fact, detunings bigger than
$\delta_0$ induce revivals of entanglement (upper right) and discord
(lower left and right). The entanglement dynamics of the upper panel
allows us to point out an important issue: $\delta>\delta_0$ is a
necessary condition for an oscillating $\beta(t)$, though revivals of
entanglement also depend on the rescaled coupling $\lambda$. In other
words, when $\delta > \delta_0$, the symplectic eigenvalue
$\tilde{d_{-}}$ of (\ref{eq:entanglement}) flows in time in unison with
$\beta(t)$, without necessarily violating the separability condition
$\tilde{d_{-}} \geq \frac12$. This explains the presence of a plateau in
the entanglement of the common scenario with $\delta = \delta_0$. 
\par
\begin{figure}[h]
\includegraphics[width=0.48 \columnwidth]{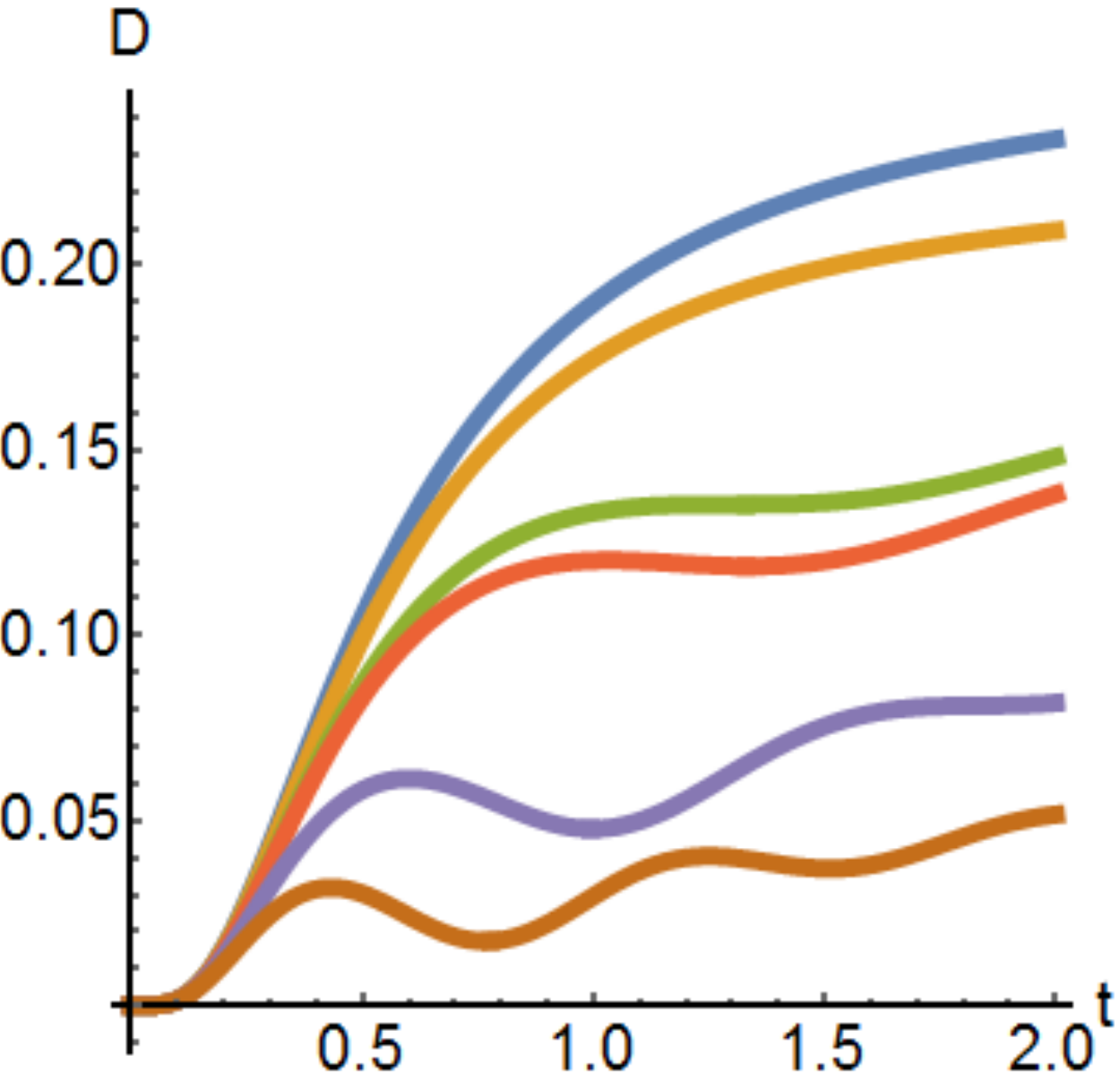}
\includegraphics[width=0.48 \columnwidth]{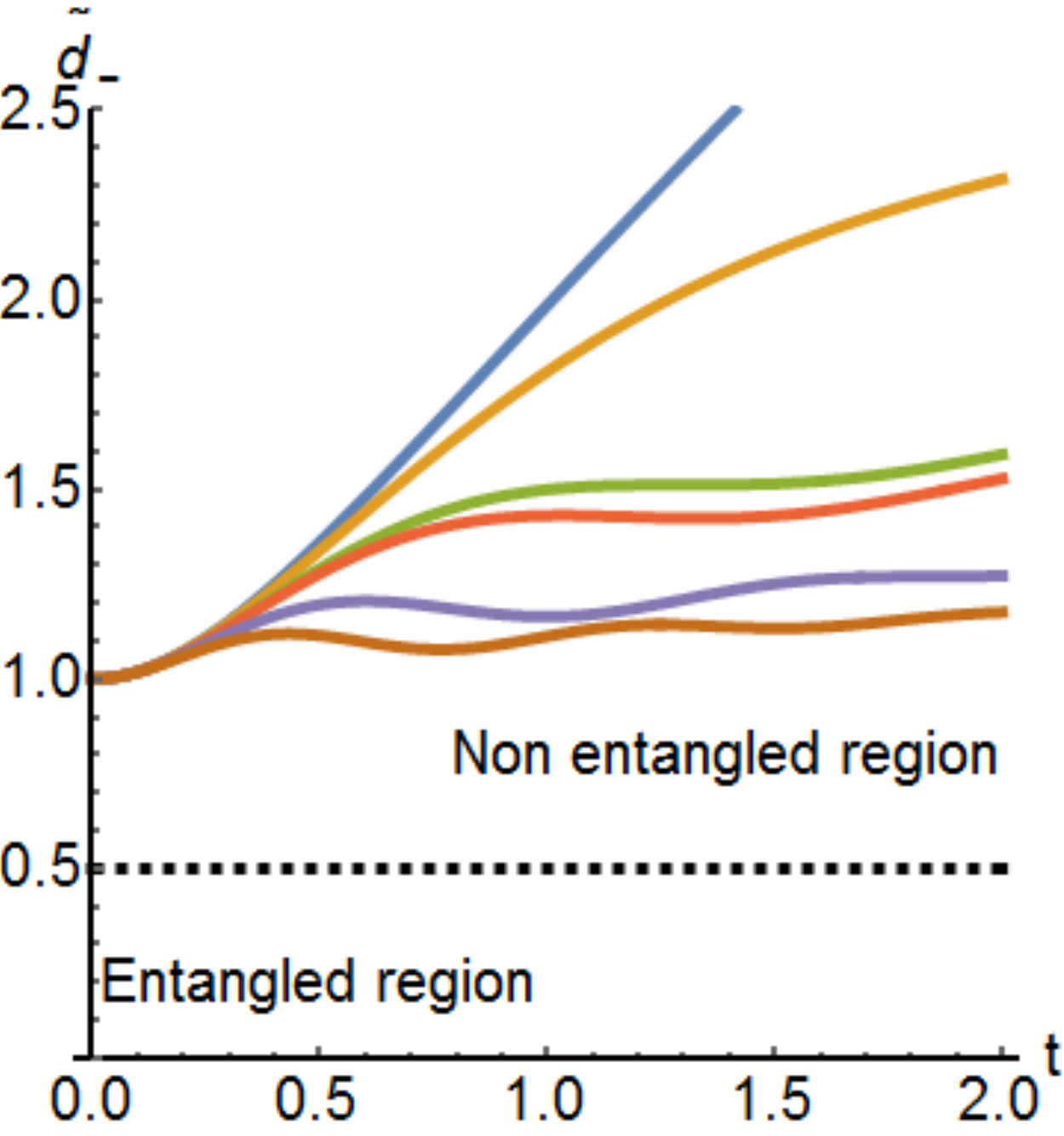}
\includegraphics[width=0.43 \columnwidth]{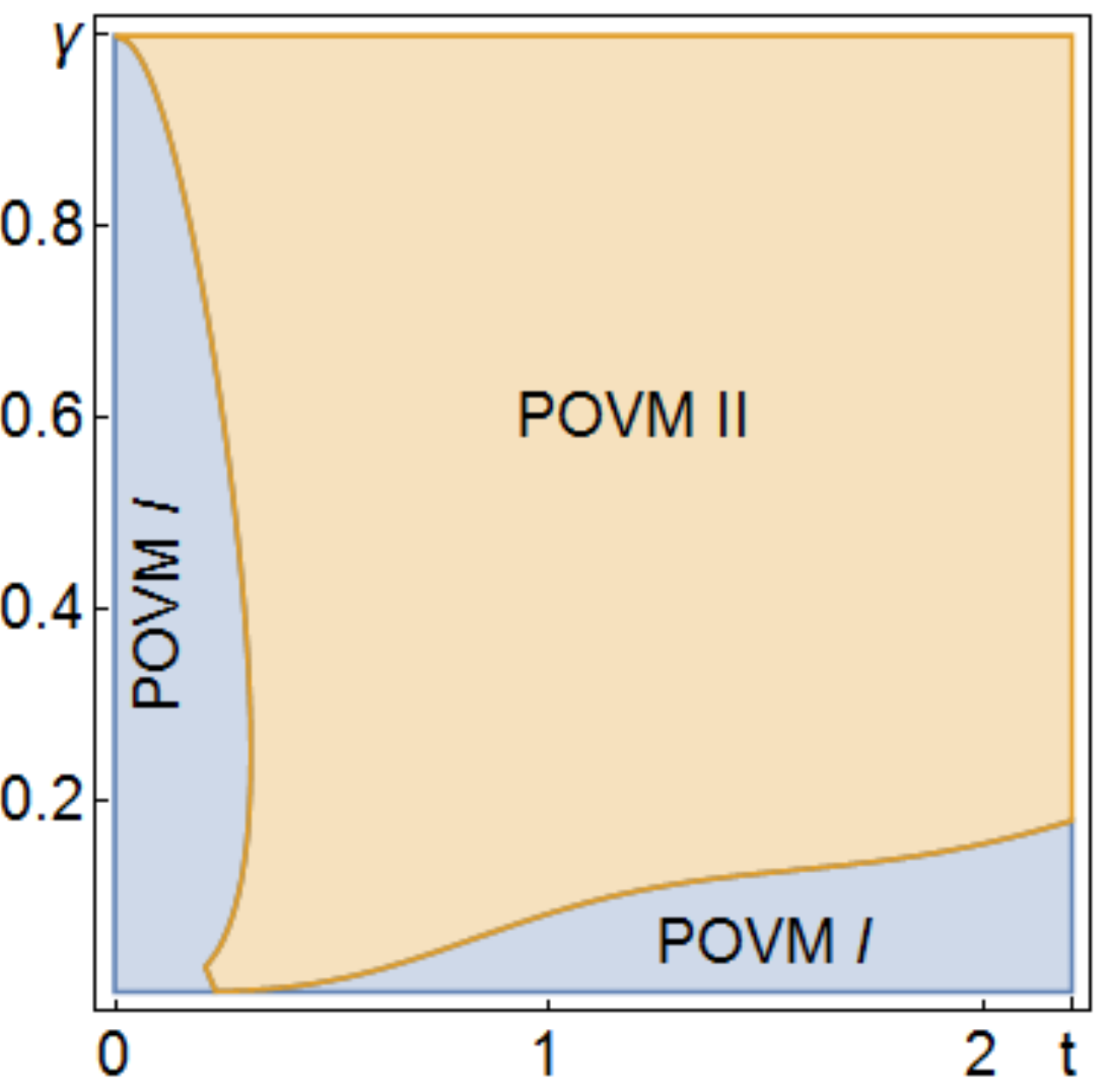}
\includegraphics[width=0.50 \columnwidth]{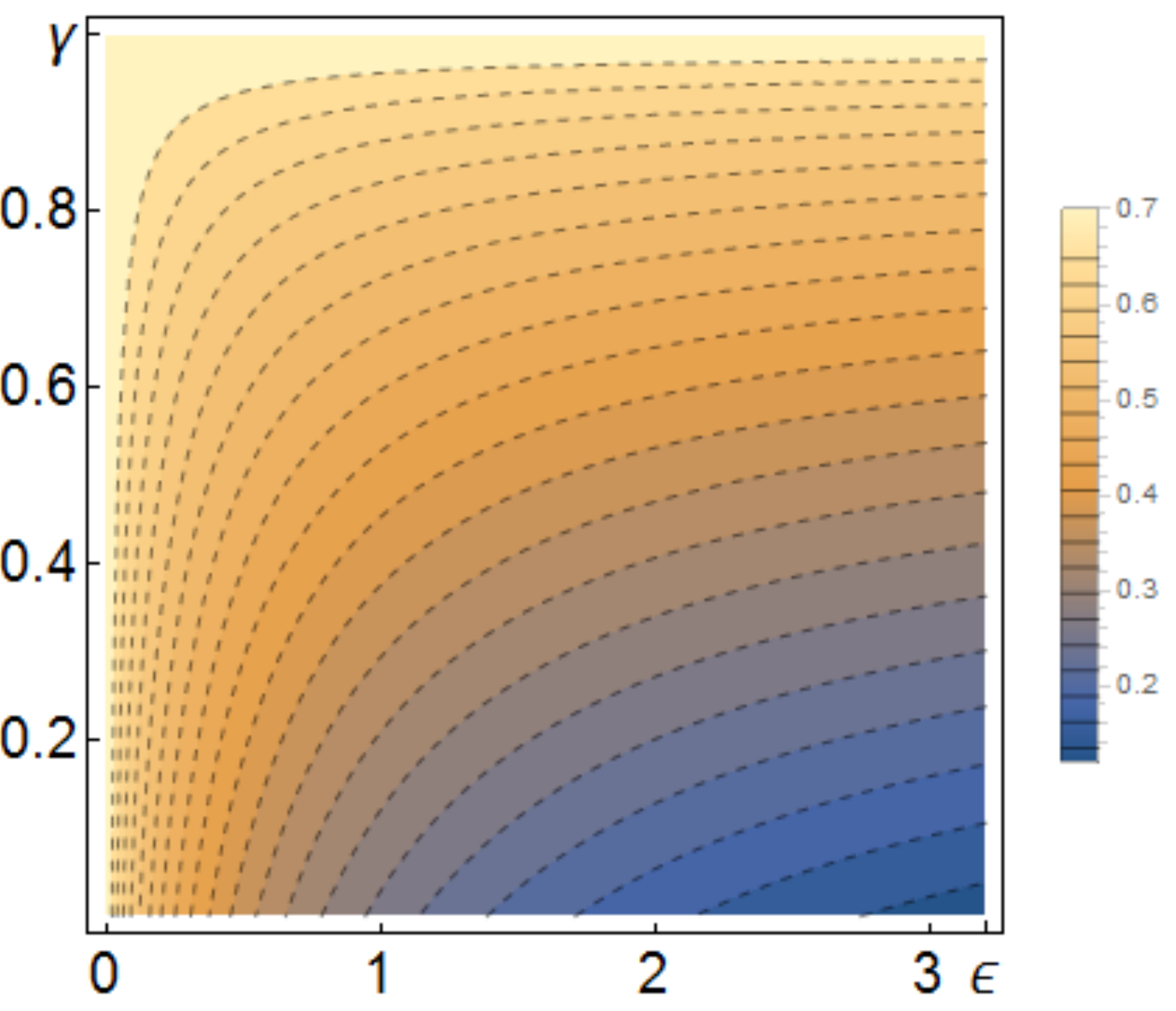}
\caption{Upper panels: Correlations of a two-mode thermal state. Left panel: discord dynamics in time $t$ for different values of $\delta$. The initially zero-discord state becomes a discord state because of the interaction. Right panel: dynamics of symplectic eigenvalue $\tilde d_{-}$ for different values of $\delta$. The state always remains separable, though becoming a discord state. In both panels we have $\epsilon =1, \gamma =0, \lambda=1$ and, from top to bottom, $\delta = 0$ (blue line), $\delta = 2$ (yellow line), $\delta = \delta_0$ (green line), $\delta = 4$ (red line), $\delta = 6$ (purple line), $\delta = 8$ (brown line). Lower left panel: Regionplot of the POVM minimizing the quantum discord. We set $\epsilon =1, \delta = 3, \lambda = 1$. Lower right panel: contourplot of the asymptotic value of the quantum discord as a function the input state parameters $\epsilon$ and $\gamma$. We set $\delta = 3, \lambda =1$.  }
\label{f:thermal}
\end{figure}

Let us now focus on the discord dynamics (see the the lower panels in
Fig. \ref{fig:ent1}).
While the entanglement shows a vanishing behavior in both scenarios in
any setup of parameters, the same cannot be said for the quantum
discord. While in the local scenario the initial discord tends to
vanish, the common interaction introduces some correlations which
clearly arise after the drop of the initial discord \cite{cic13}.  
The effect of the
common stochastic field on the dynamics of the quantum discord is even
clearer in the case of thermal input states (squeezing parameter $\gamma
=0$).  The upper left panel of Fig. \ref{f:thermal} shows the discord
evolution of the state $\rho = \nu_1 \otimes \nu_2$ in the common
scenario. 
The interaction transforms the initial zero-discord state into a discord state without affecting the separability of the input state (the symplectic eigenvalue $\tilde d_{-}$ always satisfies the condition
$\tilde d_{-} \geq \frac12$, as is apparent from the upper right panel of Fig. \ref{f:thermal}).
Furthermore, the quantum discord tends to an asymptotic value which depends on both the energy $\epsilon$ and the squeezing parameter $\gamma$ of the input Gaussian state, but it is not affected by the parameters of the environment $\lambda$ and $\delta$. Indeed, this can be seen as a consequence of the non-markovianity of the quantum map, as the long-time dynamics is influenced by the input state. A contourplot of the asymptotic value of the Discord as a function of $\epsilon$ and $\gamma$ is shown in the lower right panel of Fig. \ref{f:thermal}. Finally, we want to mention that the POVM that minimizes the quantum discord changes in time preserving the continuity of the discord itself. As an example, we report one particular scenario in the lower left panel of Fig. \ref{f:thermal} where the regions corresponding to the two POVMs are coloured differently.

\section{Non Divisibility vs Information Backflow}
\label{s:nonmark}
The presence of revivals of correlations might be interpreted as a
signature of some form of information backflow between the system and
the environment, a phenomenon typically associated to non-Markovian
effects.  It is the purpose of this section to reveal the non-Markovian
character of the Gaussian maps of both scenarios and explore the link
between non-divisibility and information backflow, analyzing the
evolution of the Fidelity of two input states.  
\par
A completely positive map $\mathcal{E}_{(t,t_0)}$ describing
the dynamical evolution of a quantum system is said to be 
{\em divisible} if it satisfies the decomposition rule 
$\mathcal{E}_{(t_2,t_0)}= \mathcal{E}_{(t_2,t_1)} 
\mathcal{E}_{(t_1,t_0)}$ for any $t_2 \geq t_1
\geq t_0$. Divisibility is often assumed to be 
the key concept to characterize non-Markovianity in the quantum regime
and a completely positive map is said to be non-Markovian it 
it violates the decomposition rule for some set of times.  
\par
In our system, it is straightforward 
to prove that the Gaussian maps (\ref{locdyn}) and (\ref{comdyn}) do not
satisfy the divisibility conditions, i.e. they cannot describe a Markovian
dynamics. In order to prove this results, we notice that the 
composition of maps $\mathcal{E}_{\tl}(\Delta t_2)
\mathcal{E}_{\tl}(\Delta t_1)$ corresponds to a convolution, 
leading to 
$\mathcal{E}_{\tl}(\Delta t_2)
\mathcal{E}_{\tl}(\Delta t_1)=\mathcal{E}_{\tl} 
(\Delta t_1 +\Delta t_2)$. This condition is satisfied 
if and only if \be
\beta(\Delta t_1 + \Delta t_2) = \beta(\Delta t_1) +\beta(\Delta t_2).
\ee
which is not satisfied for any
choice of the parameters $\delta$ and $\lambda$, thus implying that
the map is always non-Markovian. A similar proof can be obtained for
the common noise map $\mathcal{E}_{\tc}(\Delta t)$.  
\par 
We are now ready to discuss the
connections between revivals of correlations, non-divisibility and
information backflow.  As we already mentioned in the introduction,
non-Markovianity may be revealed by some witnesses, as the $BLP$ measure
or the analogue measure based on fidelity for CV systems. Both
techniques are based on the contractive property (valid for
Markovian dynamics) of the trace distance and the Bures distance,
respectively. Therefore, a non-monotonous behaviour of the trace distance
or the fidelity is a signature of non-Markovianity. Furthermore, both
these witnesses possess physical meaning: the trace distance is directly
related to the probability of discriminating two states in time, whereas 
the Bures distance may be used to evaluate upper and lower bounds of the
very same error probability defined by the trace distance. Therefore, a
non-monotonous dynamics also implies a partial clawback of
distinguishability of two input states, which has been interpreted as a
sign of a backflow of information\cite{pii03}. A measure of 
non-Markovianity $\mathcal{N}_\mathcal{F}$
can be constructed by the violation of the contractive property of the fidelity,
\begin{eqnarray}
\mathcal{N}_{\mathcal{F}} =
\frac12 \int 
\frac12 \int \left(\Big|\frac{d}{dt} D_B(\rho_1, \rho_2) \Big|+ \frac{d}{dt} D_B(\rho_1, \rho_2)  \right) dt,
\end{eqnarray}
where we used the Bures distance
\be
D_B(\rho_1, \rho_2) = \sqrt{2 \Big[1- \sqrt{\mathcal{F}(\rho_1, \rho_2)}\Big]}.
\ee
The quantity $\mathcal{N}_{\mathcal F}$ is nonzero only when the derivative of the Bures distance is positive, i.e. the contractive property is violated and the fidelity has a non-monotonous behaviour.
In our system, the fidelity between any pair of two-mode Gaussian states 
\be
\label{fidelity}
\mathcal{F}(\rho_1,\rho_2) = 
\Big[ \mbox{Tr}\sqrt{\sqrt{\rho_1} \rho_2 \sqrt{\rho_1}}\, \Big].
\ee 
may be evaluated analytically \cite{mar01}, though its expression is
cumbersome and will not reported here. 
\begin{figure}[h!]
\centering
\includegraphics[width=0.48\columnwidth]{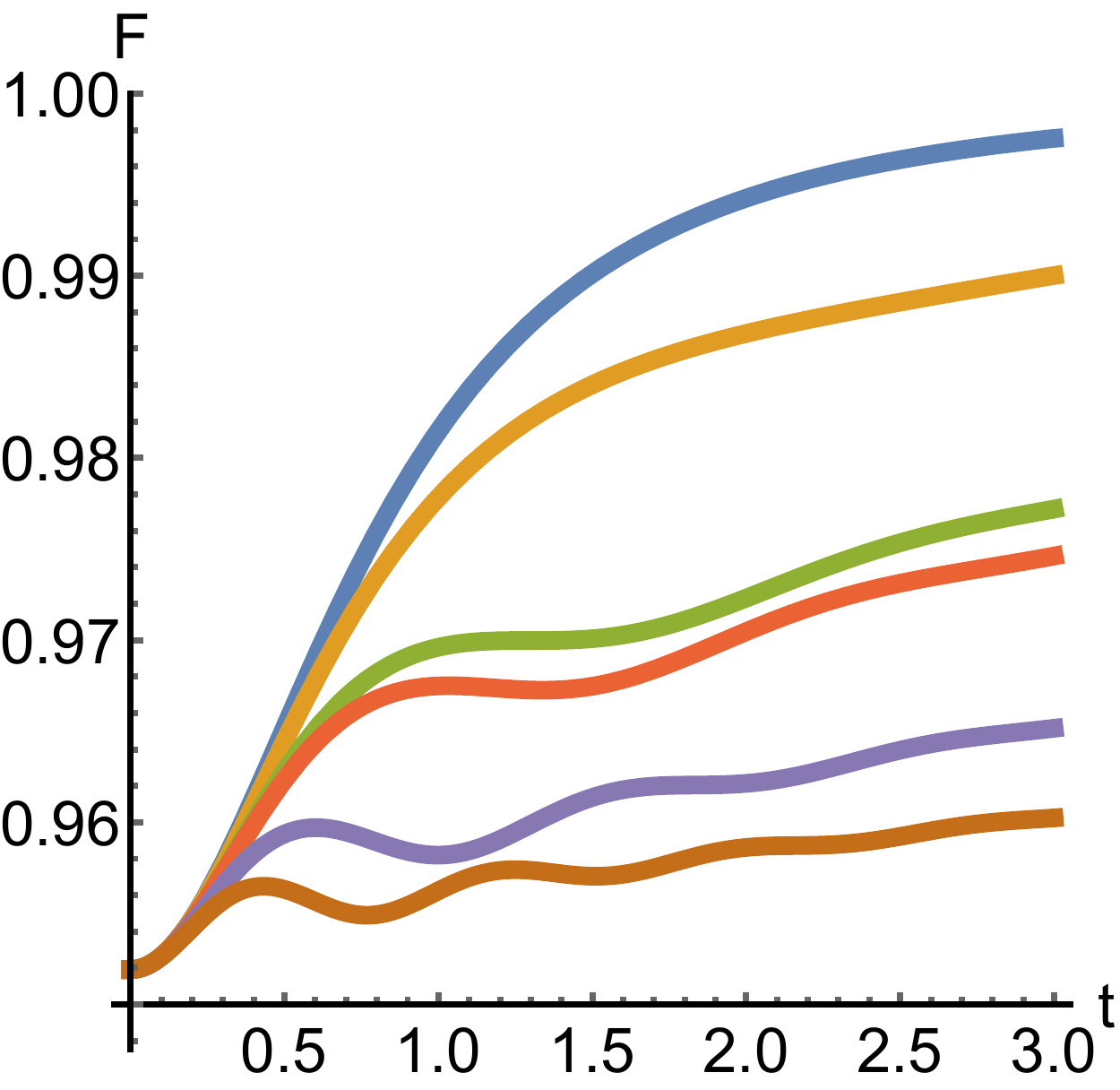}
\includegraphics[width=0.48\columnwidth]{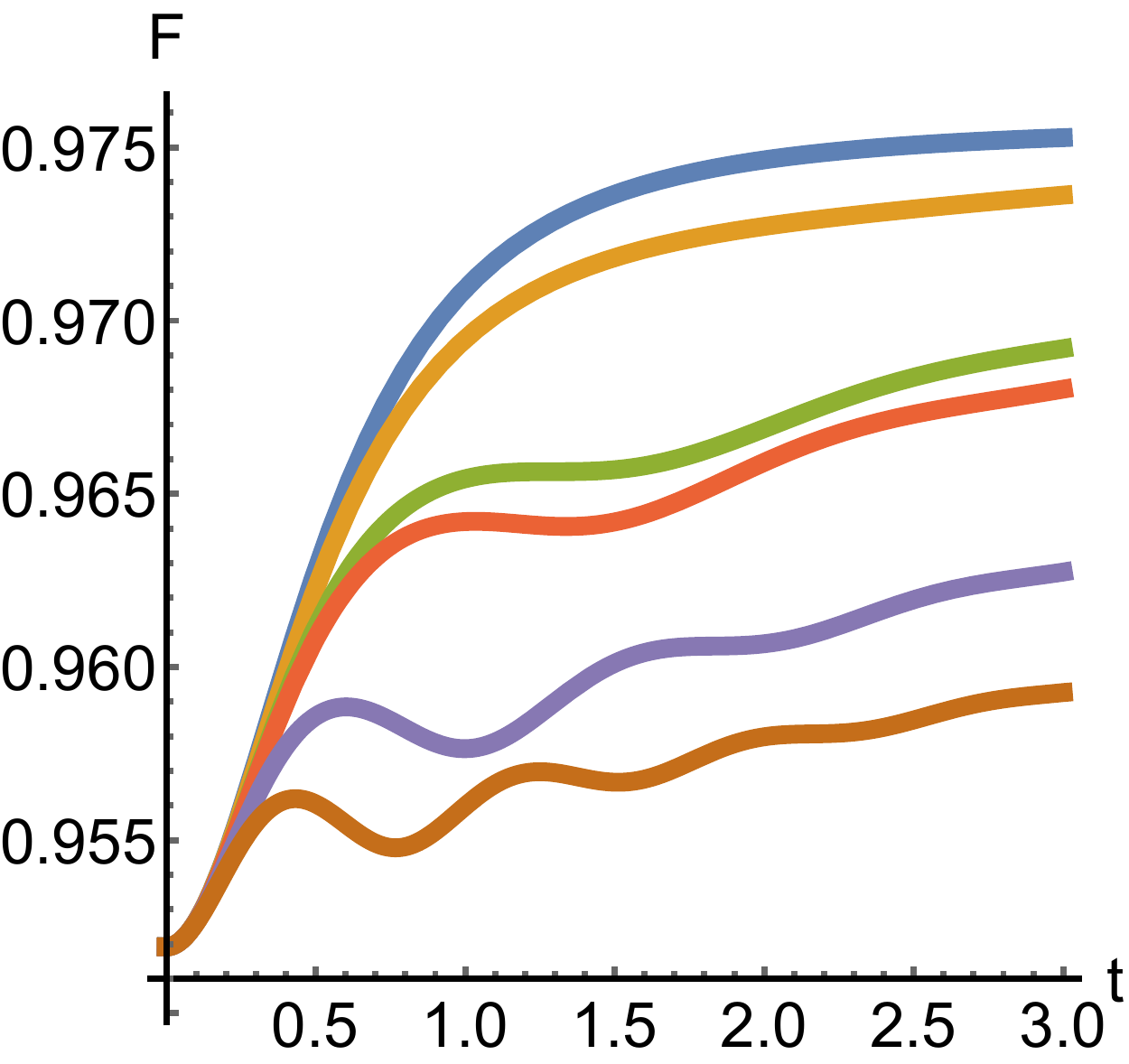}
\includegraphics[width=0.48\columnwidth]{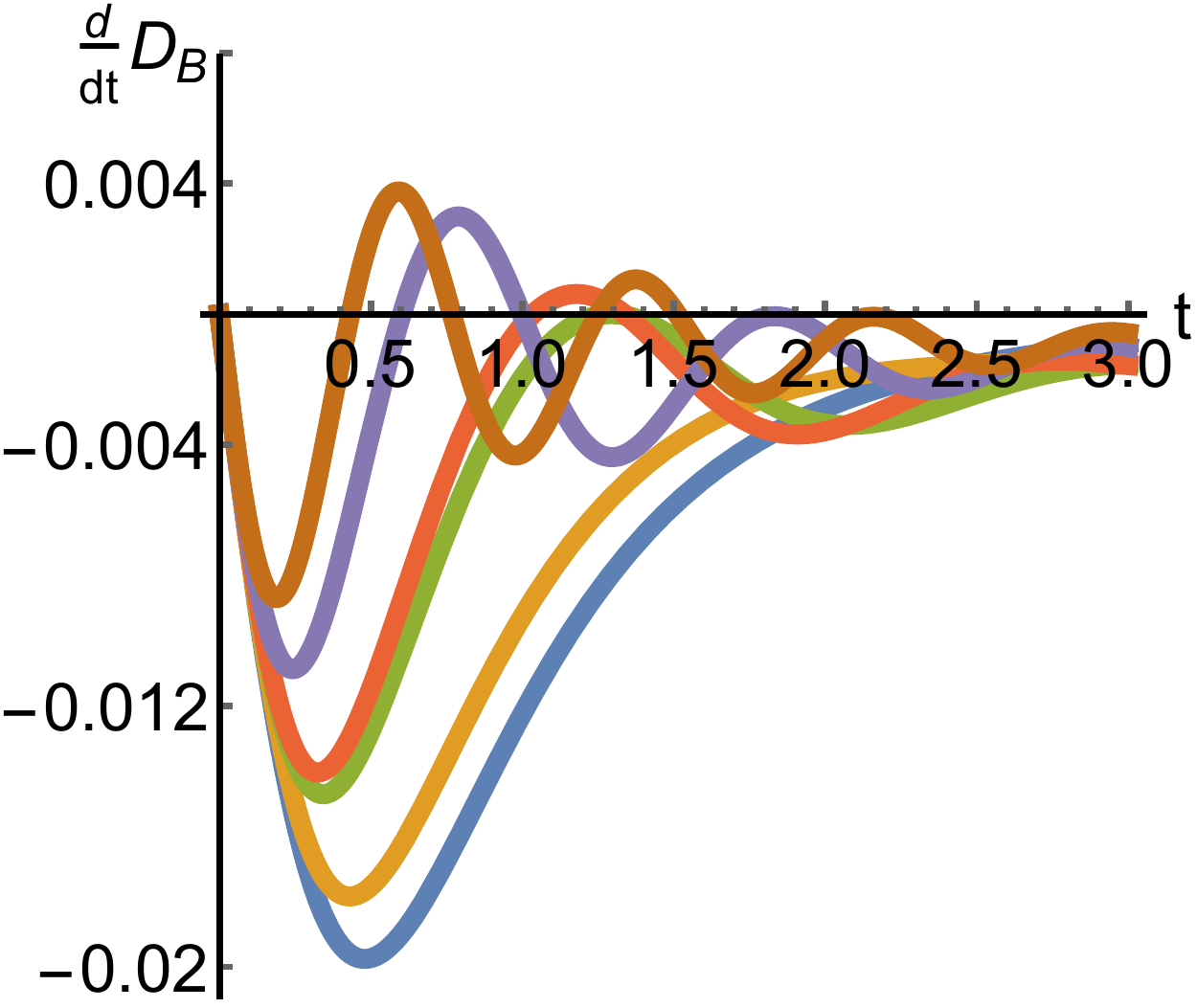}
\includegraphics[width=0.48\columnwidth]{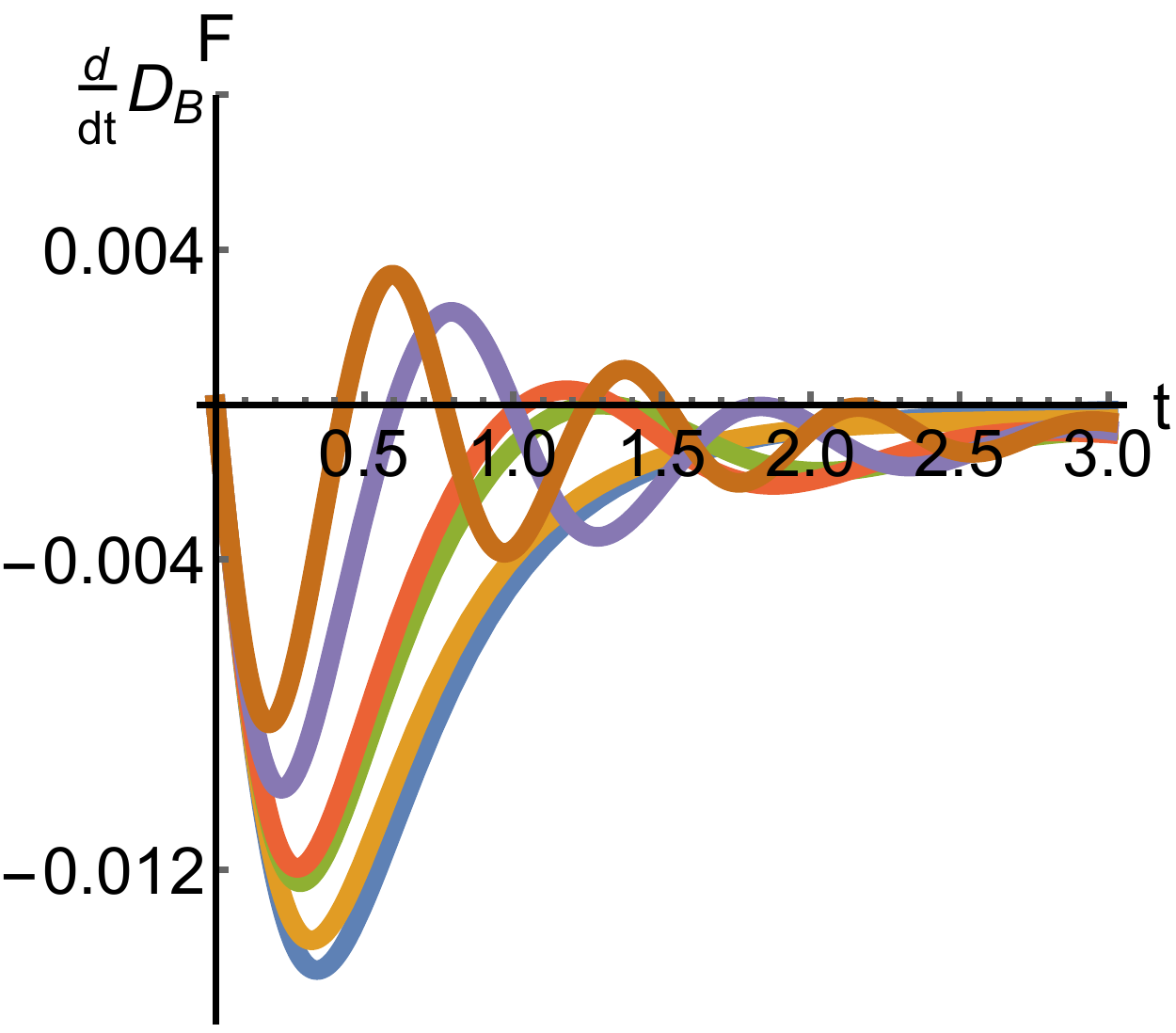}
\caption{Dynamics of Fidelity and Bures distance derivative for
different values of rescaled detuning $\delta$. Upper left panel:
dynamics of Fidelity in a local scenario.  Upper right panel: dynamics
of Fidelity in a common scenario. In both situations, Fidelity ceases to
oscillate when $\delta$ is lower than the threshold value $\delta_0$.
The distinguishability of the input states diminishes monotonically in
time and no backflow of information is detected. Lower left panel:
derivative of Bures distance in local scenario. Lower right panel:
derivative of Bures distance in common scenario. The curves with a
positive part of derivative of Bures distance have $\delta >\delta_0$
and correspond to the curves in the upper panels where oscillations of
Fidelity are shown. In all panels, we set $\epsilon_1 = 2, \epsilon_2 =
1, \gamma_1 = \gamma_2 = 1, \lambda = 1$ and, from top to bottom
$\delta = 0$ (blue line), $\delta = 2$ (yellow line), $\delta =
\delta_0$ (green line), $\delta = 4$ (red line), $\delta = 6$ (purple
line), $\delta = 8$ (brown line). } \label{f:fidelity}
\end{figure}
\par
In Fig.\ref{f:fidelity} we show the time evolution of the fidelity and the derivative of
the Bures distance
between a pair of two-mode squeezed vacuum states ($\gamma = 1$) with
different energies ($\epsilon_1 \neq \epsilon_2$).  The existence of
sets of parameters leading to a non-monotonous behavior of the fidelity
and a region of positive derivative of Bures distance
is enough to confirm the already proven non-Markovianity of both maps.
However, non-Markovianity is not detected when $\delta \leq \delta_0$,
where $\delta_0$ is the very same threshold obtained in the previous 
Section, i.e. the threshold to observe revivals of correlations. 
The same behaviour is observed for any choice of the involved
parameters, confirming that revivals of correlations are connected to
the backflow of information revealed by the fidelity measure, rather
than a feature related to non-divisibility of the map 
itself. 
\section{Conclusions}
\label{s:conclusions}
In conclusions, we have investigated the evolution of entanglement and
quantum discord for two harmonic oscillators interacting with classical
stochastic fields. We analyzed two different regimes: in the first the
two modes interact with two separate environments describing local
noise, whereas in the second case we consider a single environment
describing a situation where the two oscillators are exposed to a common
source of noise). 
\par
We have obtained the analytic form of the quantum map for both the local
and the common noise model and analyzed the dynamics of quantum 
correlations for initial states
ranging from maximally entangled to zero discord states. 
Our results show that the interaction with a classical 
environment always induces a loss
of entanglement, while the quantum discord shows a vanishing behavior in
the local scenario but may exhibit a non zero asymptotic value in the common
scenario, independently on the the initial value of the discord.
We have also shown that the interaction with a common environment is, 
in general, less detrimental than the interaction with separate ones. 
\par
Finally, we have proved the non-divisibility of the maps and found 
some structural boundaries on the existence of revivals
of correlations in terms of a threshold value of the detuning 
between the 
natural frequency of the system and the central frequency of the 
noise. The same threshold
determines the presence of backflow of information, associated to
oscillations of the fidelity between a pair of initial states. 
Overall, this suggests that non-divisibility in itself is not a 
resource to preserve quantum correlations in  our system, i.e. it
is not sufficient to observe recoherence phenomena. 
Rather, it represents a necessary prerequisite to
obtain backflow of information, which is the true ingredient to obtain
revivals of quantumness and, in turn, the physically relevant resource.
In this framework, our findings support some recent results 
\cite{hp1,hp2} about the definition of quantum non-Markovianity, 
which emphasize the fundamental role of information backflow, as opposed to
divisibility, as a key concept for the characterization of non-Markovian
dynamics in the quantum regime.
\section*{Acknowledgments}
This work has been supported by EU through the 
Collaborative Projects QuProCS (Grant Agreement 641277) and by 
UniMI through the H2020 Transition Grant 14-6-3008000-625.
The authors thanks H.-P. Breuer and B. Vacchini for discussions.


\begin{thebibliography}{99}
\bibitem{zur01} W. H. Zurek "Decoherence and the Transition from Quantum to Classical" , Phys. Today {\bf{44}} (10), 36 (1991).
\bibitem{paz01} J. P. Paz, S. Habib, and W. H. Zurek "Reduction of the wave packet: Preferred observable and decoherence time scale", Phys. Rev. D  {\bf{47}}, 488 (1993).
\bibitem{liu02} B.-H. Liu, L. Li, Y.-F. Huang, C.-F. Li, G.-C. Guo, E.-M. Laine, H.-P. Breuer, J. Piilo "Experimental control of the transition from Markovian to non-Markovian dynamics of open quantum systems", Nat. Phys. {\bf{7}}, 931 (2011).
\bibitem{smi01} A. Smirne, D. Brivio, S. Cialdi, B. Vacchini, and M. G. A. Paris "Experimental investigation of initial system-environment correlations via trace-distance evolution", Phys. Rev. A {\bf{84}}, 032112 (2011).
\bibitem{pii04} J. Piilo, S.Maniscalco,K. Harkonen, and K.-A. Suominen "Non-Markovian Quantum Jumps", Phys. Rev. Lett. {\bf{100}}, 180402 (2008).
\bibitem{smi02} A. Smirne, S. Cialdi, G. Anelli, M. G. A. Paris, and B. Vacchini "Quantum probes to experimentally assess correlations in a composite system", Phys. Rev. A {\bf{88}}, 012108 (2013).
\bibitem{gar01} C.W. Gardiner, {\it{Harndbook of Stochastic Methods}}, (Springer, Berlin, 1983).
\bibitem{hel01} J. Helm and W. T. Strunz "Quantum decoherence of two qubits", Phys. Rev. A {\bf{80}}, 042108 (2009).
\bibitem{hel02} J. Helm, W. T. Strunz, S. Rietzler, and L. E. W\"urflinger "Characterization of decoherence from an environmental perspective", Phys. Rev. A {\bf{83}}, 042103 (2011).
\bibitem{cro01} D. Crow and R. Joynt "Classical simulation of quantum dephasing and depolarizing noise", Phys. Rev. A {\bf{89}}, 042123 (2014).
\bibitem{wit01} W. M. Witzel, K. Young, and S. Das Sarma "Converting a real quantum spin bath to an effective classical noise acting on a central spin", Phys. Rev. B {\bf{90}}, 115431 (2014).
\bibitem{str02}W. T. Strunz, L. D\'iosi, and N. Gisin "Open System Dynamics with Non-Markovian Quantum Trajectories", Phys. Rev. Lett. {\bf{82}}, 1801 (1999).
\bibitem{sto01}J. T. Stockburger and H. Grabert "Exact c-Number Representation of Non-Markovian Quantum Dissipation", Phys. Rev. Lett. {\bf{88}}, 170407 (2002).
\bibitem{ast01}  O. Astafiev, Yu. A. Pashkin, Y. Nakamura, T. Yamamoto, and J. S. Tsai "Quantum Noise in the Josephson Charge Qubit", Phys. Rev. Lett. {\bf{93}}, 267007 (2004).
\bibitem{gal01} Y. M. Galperin, B. L. Altshuler, J. Bergli, and D. V. Shantsev "Non-Gaussian Low-Frequency Noise as a Source of Qubit Decoherence", Phys. Rev. Lett. {\bf{96}}, 097009 (2006).
\bibitem{abe01} B. Abel and F. Marquardt "Decoherence by quantum telegraph noise: A numerical evaluation", Phys. Rev. B {\bf{78}}, 201302(R) (2008).
\bibitem{str01} T. Grotz, L. Heaney and W.T. Strunz "Quantum dynamics in fluctuating traps: Master equation, decoherence, and heating", Phys. Rev. A, {\bf{74}}, 022102 (2006).
\bibitem{tur01} Q.A. Turchette, C. J. Hyatt, B.E. King, C. A. Sackett, D. Kielpinski, W. M. Itano, C. Monroe and D. J. Wineland "Decoherence and decay of motional quantum states of a trapped atom coupled to engineered reservoirs", Phys. Rev. A {\bf{62}}, 053807 (2000).
\bibitem{bel02} B. Bellomo, R. Lo Franco and G. Compagno "Entanglement dynamics of two independent qubits in environments with and without memory", Phys. Rev. Lett. {\bf{77}}, 032342 (2008).
\bibitem{maz02} L. Mazzola, S. Maniscalco, J. Piilo, K.-A. Suominen, and B. M. Garraway "Sudden death and sudden birth of entanglement in common structured reservoirs", Phys. Rev. A {\bf{79}}, 042302 (2009).

\bibitem{rhp} A. Rivas, S. F. Huelga, and M. B. Plenio, "Entanglement and Non-Markovianity of Quantum Evolutions", Phys. Rev. Lett. {\bf{105}}, 050403 (2010).

\bibitem{ill01} G. Torre, W. Roga and F. Illuminati, "Non-Markovianity of Gaussian Channels", Phys. Rev. Lett. {\bf{115}}, 070401 (2015).

\bibitem{ade01}
L. A. M. Souza, H. S. Dhar, M. N. Bera, P. Liuzzo-Scorpo and G. Adesso, "Gaussian interferometric power as a measure of continuous-variable non-Markovianity",
Phys. Rev. A {\bf{92}}, 052122 (2015).

\bibitem{pii01} H.-P. Breuer, E.-M. Laine and J. Piilo "Measure for the Degree of Non-Markovian Behavior of Quantum Processes in Open Systems", Phys. Rev. Lett. {\bf{103}}, 210401 (2009).

\bibitem{nmc11} 
{R. Vasile, S. Maniscalco, M. G. A. Paris, H.-P. Breuer, J. Piilo "Quantifying non-Markovianity of continuous-variable Gaussian dynamical maps"}, 
Phys. Rev. A {\bf 84}, 052118 (2011).

\bibitem{mag1} W. Magnus "On the exponential solution of differential equations for a linear operator", Comm. Pure and Appl. Math. {\bf{7}}, 649 (1954).
\bibitem{mag2} S. Blanes, F. Casas, J.A. Oteo, and J. Ros "The Magnus expansion and some of its applications", 
Phys. Rep. {\bf{470}}, 151 (2008).


\bibitem{liu01} J.Liu, X.-M. Lu and X. Wang "Nonunital non-Markovianity of quantum dynamics", Phys. Rev. A  {\bf{87}} 042103 (2013).
\bibitem{hai01} P.Haikka, J.D. Cresser and S. Maniscalco "Comparing different non-Markovianity measures in a driven qubit system", Phys. Rev. A, {\bf{83}} 012112 (2011).
\bibitem{pii02} L. Mazzola, E.-M. Laine,H.-P. Breuer, S.Maniscalco and J. Piilo"Phenomenological memory-kernel master equations and time-dependent Markovian processes", Phys. Rev. A {\bf{81}}, 062120 (2010).

\bibitem{hp1} 
S. Wissmann, B. Vacchini and H.-P. Breuer, "Generalized trace 
distance measure connecting quantum and
classical non-Markovianity", Phys. Rev. A {\bf 92}, 042108 (2015).

\bibitem{bel01} B. Bellomo, R. Lo Franco and G. Compagno "Non-Markovian Effects on the Dynamics of Entanglement", Phys. Rev. Lett. {\bf{99}}, 160502 (2007).



\bibitem{ben01} C. Benedetti, M.G.A. Paris and S. Maniscalco "Non-Markovianity of colored noisy channels", Phys. Rev. A {\bf{89}}, 012114 (2014).
\bibitem{rlf01} R. Lo Franco, B. Bellomo, E. Andersson and G. Compagno "Revival of quantum correlations without system-environment back-action", Phys. Rev. A {\bf{85}}, 032318 (2012).
\bibitem{catstf} J. Trapani, M. Bina, S. Maniscalco, M. G. A. Paris "Collapse and revival of quantum coherence for a harmonic oscillator interacting with a classical fluctuating environment",
Phys. Rev. A \textbf{91}, 022113 (2015).

\bibitem{sim1} R. Simon "Peres-Horodecki Separability Criterion for Continuous Variable Systems", Phys. Rev. Lett. {\bf{84}} 2726 (2000).

\bibitem{gio0} P. Giorda, M. G. A. Paris "Gaussian Quantum Discord", 
Phys. Rev. Lett. \textbf{105}, 020503 (2010).
\bibitem{gio1} G. Adesso and A. Datta "Quantum versus Classical Correlations in Gaussian States", Phys. Rev. Lett. {\bf{105}},
030501 (2010).  

\bibitem{hmqd}
{R. Blandino, M. G. Genoni, J. Etesse, M. Barbieri, M. G. A. Paris, P.
Grangier, R. Tualle-Brouri "Homodyne Estimation of Gaussian Quantum Discord"}, Phys. Rev. Lett \textbf{109}, 180402 (2012).

\bibitem{bru15}
M. Brunelli, C. Benedetti, S. Olivares, A. Ferraro, M. G. A. Paris "Single- and two-mode quantumness at a beam splitter",
Phys. Rev. A, Phys. Rev. A \textbf{91}, 062315 (2015).

\bibitem{pir14}
S. Pirandola, G. Spedalieri, S. L. Braunstein, N. J. Cerf, S. Lloyd "Optimality of Gaussian Discord", 
Phys. Rev. Lett.{\bf 113}, 140405 (2014)

\bibitem{cic13}
F. Ciccarello, V. Giovannetti "Creating quantum correlations through local nonunitary memoryless channels", Phys. Rev. A {\bf 85}, 010102(R)
(2012)
\bibitem{pii03} E.-M. Laine, J. Piilo and H.-P. Breuer "Measure for the non-Markovianity of quantum processes",
Phys. Rev. A {\bf{81}}, 062115 (2010).
\bibitem{mar01} P. Marian, T. A. Marian "Uhlmann fidelity between two-mode Gaussian states", Phys. Rev. A {\bf{86}}, 022340 (2012).

\bibitem{hp2} 
H.-P. Breuer, E.-M. Laine, J. Piilo and B. Vacchini, 
"Non-Markovian dynamics in open quantum systems", preprint 
arXiv:1505.01385 (2015).





\end{thebibliography}
\end{document}